\documentclass[aps,showpacs,twocolumn,eqsecnum,floatfix,amsmath,amssymb]{revtex4}

\usepackage{epsfig}
\usepackage{latexsym}
\usepackage{bm,latexsym}
\usepackage{mathrsfs}

\begin{document}


\title{Hyperbolicity of scalar-tensor theories of gravity}

\author{Marcelo Salgado$^1$}
\email{marcelo@nucleares.unam.mx}

\author{David Mart\'\i nez-del R\'\i o$^1$}
\email{david.martinez@nucleares.unam.mx}

\author{Miguel Alcubierre$^1$}
\email{malcubi@nucleares.unam.mx}

\author{Dar\'\i o N\'u\~nez$^{1,2}$
}
\email{nunez@nucleares.unam.mx}

\affiliation{$^1$Instituto de Ciencias Nucleares, Universidad Nacional
Aut\'onoma de M\'exico, A.P. 70-543, M\'exico D.F. 04510, M\'exico \\
$^2$Max-Planck Institut f\"ur Gravitationsphysik, 
Albert Einstein Institut, 14476 Golm, Germany}


\date{\today}


\begin{abstract}
Two first order strongly hyperbolic formulations of scalar-tensor
theories of gravity allowing nonminimal couplings (Jordan frame) are
presented along the lines of the 3+1 decomposition of spacetime. One
is based on the Bona-Mass\'o formulation, while the other one employs a
conformal decomposition similar to that of
Baumgarte-Shapiro-Shibata-Nakamura. A modified Bona-Mass\'o slicing
condition adapted to the scalar-tensor theory is proposed for the
analysis. This study confirms that the scalar-tensor theory has a well
posed Cauchy problem even when formulated in the Jordan frame.
\end{abstract}


\pacs{
04.50.-h, 
04.20.Ex, 
04.25.D-, 
95.30.Sf  
}


\maketitle


\section{Introduction}
\label{sec:introduction}

Scalar-tensor theories of gravity (STT) are alternative theories of
gravitation where a scalar field is coupled nonminimally to the
curvature associated with the {\em physical metric}\/ (this is
the so-called {\em Jordan frame}\/ representation).  The term
``physical metric'' refers to a situation where test particles follow
the geodesics of that metric.  The variation of the action of the STT
with respect to the physical metric gives rise to field equations which contain an
effective energy-momentum tensor (EMT) involving second order
derivatives in time and space of the scalar field.  Such EMT has the
property that ``ordinary matter'', {\em i.e.} matter associated with
fields other that the scalar field, obeys the (weak) equivalence
principle which mathematically translates into a conserved EMT for
ordinary matter alone.

Since {\em a priori}\/ it was not clear how such second order
derivatives could be eliminated (in terms of lower order derivatives),
or managed so as to obtain a quasilinear system of hyperbolic
equations for which the Cauchy problem was well-posed (in the Hadamard
sense), many people decided to abandon this approach in favor of the
so-called {\em Einstein frame}\/ representation where the nonminimal
coupling (NMC) is absorbed into the curvature by means of a conformal
transformation of the metric.  The new conformal metric is unphysical
in the sense that (non-null) test particles do not follow the
geodesics of that metric.  However, the mathematical advantage is that
the field equations for the nonphysical metric resemble the
standard Einstein field equations with an unphysical effective EMT
which involves at most first order derivatives of a suitable
transformed scalar field (this EMT is unphysical because the
``ordinary matter'' part is not separately conserved). In the Einstein
frame one can show that by using standard gauges ({\em e.g.} harmonic
gauge) the field equations acquire the form required in the
application of theorems that establish the well-posedness of the
Cauchy problem.

In view of the apparent mathematical advantages and disadvantages of
the Jordan and Einstein frames, several widespread misconceptions
became common in the literature. One of these concerned the statement
that the Cauchy problem is only well-posed in the Einstein
frame~\cite{Faraoni04}. In Ref.~\cite{Salgado06}, however, one of us
showed that this is not the case by following two different
approaches.  One was in the spirit of a second order analysis
consisting on reducing the set of field equations, both for the metric
components and the scalar field, to a manifestly quasilinear diagonal
second order hyperbolic form (the ``reduced'' field equations). This
was achieved by manipulating the field equations in a way that allowed
one to express the d'Alambertian on the scalar field in terms of at
most first order derivatives, and also by implementing a modified
harmonic gauge which was adapted to the STT. Such a modified gauge
allowed one to eliminate the remaining second order derivatives of the
scalar field which had previously prevented the applicability of
Leray's theorem (see {\em e.g.} Ref.~\cite{Wald84} for the theorem).

The second approach followed in \cite{Salgado06}, and which was
directly related with the initial value problem, consisted in
recasting the field equations of the STT in a 3+1 or
Arnowitt-Deser-Misner (ADM)~\cite{Arnowitt62} form written {\em \`a
la} York~\cite{York79} (hereafter referred to as the ADMY
equations). To do so it was necessary to define suitable variables,
followed by a manipulation of the equations in order to obtain well-defined 
constraints (independent of the choice of the lapse and
shift), as well as first order in time evolution equations for the
extrinsic curvature and the scalar-field variables.  In
Ref.~\cite{Salgado06} the main goal was to focus in the 3+1 approach
rather than the second order one because of the subsequent numerical
applications we had in mind. In fact, almost all the modern codes used
in numerical relativity are based on first order in time formulations,
and therefore we wanted to adapt them for the analysis of phenomena in
STT.

Now, when taking the limit of pure general relativity (GR) ({\em i.e.} no no-minimal
coupling), the 3+1 equations presented in~\cite{Salgado06} reduce to
the standard ADMY equations with a minimally coupled scalar field plus
ordinary matter sources. It is well-known that the ADMY equations are
not strongly hyperbolic (see~\cite{Alcubierre08a} for a detailed
discussion), and therefore the corresponding equations for STT
described in~\cite{Salgado06} were expected not to be strongly
hyperbolic either. As already mentioned in~\cite{Salgado06}, such
equations were only the first step towards a first order strongly
hyperbolic system for which the well-posedness of the Cauchy problem
could be established.

The aim of this paper is then to fill that gap, and to obtain a first
order strongly hyperbolic system of partial differential equations
based on the 3+1 equations of~\cite{Salgado06}. Actually, we will show
here two such systems: one that is related to the
Bona-Mass\'o-Seidel-Stela (BMSS) formulation~\cite{Bona94b} (which in
turn is based on the Bona-Mass\'o (BM) formulation~\cite{Bona92}), and a
second one based on the Baumgarte-Shapiro-Shibata-Nakamura (BSSN)
conformal decomposition~\cite{Baumgarte:1998te,Shibata95}.  As was
done in~\cite{Salgado06}, a modified BM slicing condition will be used
for the STT, while the shift vector will be taken as an {\em a
priori}\/ given function of the coordinates.

A last comment about the Jordan vs. Einstein frame is in
order. In~\cite{Salgado06} it was a matter of principle to show that
the Cauchy problem could be well formulated in the Jordan frame. It
then became clear that there was no fundamental reason to continue
using the Einstein frame. After all, the Jordan frame is the one
associated with the (physical) quantities that are to be confronted
with the observations. Moreover, the constraint and evolution
equations are not more involved than those of usual GR, so from a
numerical point of view the use of the Jordan frame does not add much
complexity to the analysis. On the other hand, the direct use of the
Jordan frame allows a better physical interpretation of the results
and avoids the potential problems that might arise in the Einstein
frame in cases when the conformal transformations back to the physical
metric are not well-defined.  Finally, it has been argued that some of
the energy densities associated with the Jordan frame effective EMT
(JFEMT) are not positive definite and may therefore be unphysical, and
also the corresponding ADM mass could then become negative. While it
is certainly true that the JFEMT does not satisfy the energy
conditions in general, it is very likely that any physical
configuration consistent with observations will not carry (total)
negative energy.  This is because the deviations of any alternative
theory from GR are presumably very small in order to reproduce many of
the current observations (cf. Ref.~\cite{Will93}). Actually, in any
viable STT cosmology it turns out that the total energy density of the
Universe is always positive, even when taking into account the
negative contributions due to the
NMC~\cite{Salgado96,Salgado97a,Salgado97b,Quevedo99}. Again, this is
because the scalar-field contributions (positive or negative) to the
total energy density should be in agreement with several observations
related to the past and present history of the Universe. For instance,
the expansion rate of the Universe modified by the NMC contributions
has to be consistent with the one required to produce the correct
abundance of primordial nucleosynthesis~\cite{Salgado96}. In addition,
the density perturbations in STT should also match the Cosmic
Microwave Background data. In other words, a STT cosmology which
produces a total effective negative energy density of the Universe at
any given epoch will surely not be consistent with observations. By
the same token, in a consistent STT cosmology the negative
contributions (if any) are to be naturally suppressed by the positive
ones, providing a consistent Universe.

In the case of astrophysical applications ({\em e.g.} neutron star
models) within a class of STT with a positive definite NMC function
(which implies a positive definite effective gravitational
``constant''), the ADM mass turns to be always positive despite the
negative contributions to the energy density due to the
NMC~\cite{Salgado98}. Moreover, the value of the NMC cannot be very
high as otherwise the corresponding STT would put in jeopardy the
agreement with the binary pulsar observations~\cite{Damour96}.

Of course, the nice thing about matter fields satisfying the energy
conditions is the applicability of several theorems ({\em e.g.}
positive mass theorems and singularity theorems; see~\cite{Wald84} and
references therein). However, the fact that the JFEMT does not satisfy
the energy conditions in general simply indicates that the theorems
cannot say anything about the positivity of mass or the formation of
singularities in this case. This is not a problem of physical but of
mathematical character. But again, the nonpositive definite terms
arising from the NMC will be surely bounded if the STT in hand is to
be consistent with the current observations and experiments, and
therefore the nice features of an EMT respecting the energy conditions
will also very likely appear in any observationally consistent STT. Of
course, all these arguments are only sustained by numerical
experiments when constructing viable phenomenological models (in
cosmology and compact objects), and therefore do not constitute a
theorem. It would then be quite interesting to explore the possibility
of proving positive energy and singularity theorems in the Jordan
frame if one can bound the non-positive definite contribution
associated with the JFEMT.

This paper is organized as follows. In Sec.~\ref{sec:ADM} we introduce
the STT and the 3+1 equations described in~\cite{Salgado06}. The
notation of several variables will be slightly modified relative to
Ref.~\cite{Salgado06} in order to match with the one usually employed
in numerical relativity.  In Sec.~\ref{sec:hyper} a fully first order
strongly hyperbolic system of the field equations of STT is obtained
along the lines of the BMSS and BSSN formulations of GR. We conclude
with a discussion of future numerical applications of the hyperbolic
systems presented here. Finally, in Appendix A we include the complete
set of equations for both formulations, and in Appendix B we introduce
a simple example to complement the ideas of
Sec.~\ref{sec:hyper}.


\section{Scalar-tensor theories of gravity}
\label{sec:ADM}

The general action for STT with a single scalar field is given by
\begin{eqnarray}
\label{jordan}
S[g_{ab}, \phi, {\mbox{\boldmath{$\psi$}}}] &=&
\!\! \int \!\! \left\{ \frac{F(\phi)}{16\pi G_0} R
- \frac{1}{2}(\nabla \phi)^2 - V(\phi)\right\} \! \sqrt{-g} \: d^4x
\nonumber \\
&+& S_{\rm matt}[g_{ab}, {\mbox{\boldmath{$\psi$}}}] \; ,
\end{eqnarray}
with $\phi$ the nonminimally coupled scalar field, and where
${\mbox{\boldmath{$\psi$}}}$ represents collectively the matter
fields, {\em i.e.} fields other than $\phi$, and $G_0$ is the usual
gravitational constant. (we use units such that $c=1$).

The representation of the STT given by Eq.~(\ref{jordan}) is called
the {\em Jordan frame}\/ representation. The field equations obtained
from the action~(\ref{jordan}) are given by~\footnote{Here and in what
follows, Latin indices from the first letters of the alphabet
$a,b,c,...$ are four-dimensional and run $0-3$, while Latin indices
starting from $i$ $(i,j,k,...)$ are three-dimensional and run $1-3$.}
\begin{eqnarray}
\label{Einst}
G_{ab} &=& 8\pi G_0 T_{ab}\,\,\,\,, \\
\label{KGo}
\Box \phi &+& \frac{1}{2}f^\prime R = V^\prime \,\,\,,
\end{eqnarray}
where $^\prime$ indicates $\partial_\phi$,
$\Box:=g^{ab}\nabla_a\nabla_b$ is the covariant d'Alambertian
operator, $G_{ab}=R_{ab}-\frac{1}{2}g_{ab}R$, and
\begin{eqnarray}  
\label{effTmunu}
T_{ab} &:=& \frac{G_{{\rm eff}}}{G_0}\left(\rule{0mm}{0.5cm} T_{ab}^f + 
T_{ab}^{\phi} + T_{ab}^{{\rm matt}}\right)\,\,\,\,, \\
\label{TabF}
T_{ab}^f &:= & \nabla_a\left(f^\prime 
\nabla_b\phi\right) - g_{ab}\nabla_c \left(f^\prime 
\nabla^c \phi\right) \,\,\,\,, \\
T_{ab}^{\phi} &:= & (\nabla_a \phi)(\nabla_b \phi) - g_{ab}
\left[ \frac{1}{2}(\nabla \phi)^2 + V(\phi)\right ] \,, \quad \\
\label{Geff}
G_{{\rm eff}} &:=& \frac{1}{8\pi f} \,\,\,\,,\,\,\,\,f:=\frac{F}{8\pi G_0} 
\,\,\,\,.
\end{eqnarray}

Using Eq.~(\ref{Einst}), the Ricci scalar can be expressed in terms of
the energy-momentum tensor Eq.~(\ref{effTmunu}). Equation~(\ref{KGo})
then takes the following form,
\begin{equation}
\label{KG}
{\Box \phi} = \frac{ f V^\prime - 2f^\prime V -\frac{1}{2}f^\prime
\left( 1 +  3f^{\prime\prime} 
\right)(\nabla \phi)^2 + \frac{1}{2}f^\prime T_{{\rm matt}} }
{f\left(1 + \frac{3{f^\prime}^2}{2f}\right) }\,\,\,\,,
\end{equation}
where $T_{{\rm matt}}$ stands for the trace of $T^{ab}_{{\rm matt}}$
(the subscript ``matt'' refers to matter fields other that $\phi$).

Now, the Bianchi identities imply that
\begin{equation}
\nabla _{c }T^{c a }=0\,\,\,\,.
\end{equation}
However, the use of the field equations leads, as mentioned before, to
the conservation of the EMT of the matter alone
\begin{equation}
\nabla _{c }T_{{\rm matt}}^{c a }=0\,\,\,\,,
\end{equation}
which implies the fulfillment of the (weak) equivalence principle.

In what follows we shall use the 3+1 formalism of spacetime
\cite{York79,Gourgoulhon07} in order to recast the field equations as
a Cauchy initial value problem. The following quantities turn out to be
useful in the 3+1 decomposition of the field equations of
STT~\cite{Salgado06}:
\begin{eqnarray}
\label{Q}
Q_a &:=& D_a \phi \,\,\,,\\
\label{Pidef}
\Pi &:=& {\cal L}_{\mbox{\boldmath{$n$}}} \phi \,\,\,,
\end{eqnarray}
where ${\cal L}_{\mbox{\boldmath{$n$}}}$ stands for the Lie derivative
along the normal $n^a$ to the spacelike hypersurfaces $\Sigma_t$, and
the operator $D_a$ is the covariant derivative compatible with the
3-metric $\gamma_{ab}$ induced on $\Sigma_t$.  In components we have
$n^a= (1/\alpha,-\beta^i/\alpha)$, where $\alpha$ and $\beta^i$ are
the {\em lapse function}\/ and {\em shift vector}, respectively,
associated with the spacetime coordinates $(t,x^i)$~\footnote{A
slightly different notation to that of Ref.~\cite{Salgado06} has been
used here in order to match with the one used in many references on
numerical relativity. In order to return to the notation
of~\cite{Salgado06}, one needs to perform the transformation
$\alpha \rightarrow N$, $\beta^i \rightarrow -N^i$,
$\gamma_{ij} \rightarrow h_{ij}$ for the lapse, shift and the 3-metric
respectively.}.

From definitions (\ref{Q}) and (\ref{Pidef}) it is possible to write
an evolution equation for $Q_a$:
\begin{equation}
\label{EvQ}
{\cal L}_{\mbox{\boldmath{$n$}}} Q_a =\frac{1}{\alpha} \: D_a(\alpha\Pi)
\,\,\,.
\end{equation}
On the other hand, the definition of the extrinsic curvature is given
by
\begin{equation}
\label{K_ab}
K_{ab}:= -\frac{1}{2} \: {\cal L}_{\mbox{\boldmath{$n$}}}\gamma_{ab}\,\,\,.
\end{equation}
In this way, the 3+1 decomposition of equations~(\ref{Einst}) leads to
a set of constraint equations plus the evolution equations for the
extrinsic curvature (see Ref.~\cite{Salgado06} for details).  The
Hamiltonian and momentum constraints are respectively
\begin{eqnarray}  
\label{CEHfSST}
&&
^3 R + K^2 - K_{ij} K^{ij} 
- \frac{2}{f}\left[
 f^\prime \left(\rule{0mm}{0.4cm} D_l Q^l + 
K \Pi \right) + \frac{\Pi^2}{2} \right. \nonumber \\
&& \left. + \frac{Q^2}{2}
\left(\rule{0mm}{0.4cm} 1 + 2f^{\prime\prime}\right) \right]
  = \frac{2}{f} \left[\rule{0mm}{0.4cm} E_{\rm matt} + V(\phi)\right]  \,\,\,,\\
\label{CEMfSST}
&& D_l K_{\,\,\,\,\,i}^{l} - D_i K
+\frac{1}{f}\left[\rule{0mm}{0.5cm}
f^\prime\left(\rule{0mm}{0.4cm} K_{i}^{\,\,l} Q_l + \,D_i\Pi
\right) \right. \nonumber \\
&& \left.\rule{0mm}{0.5cm}  + \Pi Q_i\left(\rule{0mm}{0.4cm} 1 + f^{\prime\prime}\right)\right] 
= \frac{1}{f} J_i^{\rm matt} \,\,\,\,,
\end{eqnarray}
where we have denoted $Q^2= Q_l Q^l$, $E_{\rm matt}:= n^a n^b
T_{ab}^{{\rm matt}}$ and $J^a_{\rm matt} := -n_d \gamma_{\,\,\,c}^a
T^{cd}_{\rm matt}$, and where $\gamma_{\,\,\,c}^a= \delta_{\,\,\,c}^a
+ n^a n_c$ is the projector onto $\Sigma_t$.  The evolution equation
for the extrinsic curvature takes the form:
\begin{eqnarray}  
\label{EDEfSST}
& & 
\!\!\!\!\!\!\!\!\!\!\!\!\!\!
\partial_t K_{\,\,\,j}^i - \beta^l \partial_l K_{\,\,\,j}^i - K_{\,\,\,l}^i
\partial_j \beta^l + K_{\,\,\,j}^l \partial_l \beta^i 
+ D^i D_j \alpha \nonumber \\
&&
\!\!\!\!\!\!\!\!\!\!\!\!\!\!
-\,^3 R_{\,\,\,j}^i \alpha - \alpha K K_{\,\,\,j}^i + \frac{\alpha}{f}\left[\rule{0mm}{0.6cm} Q^i
Q_j\left(\rule{0mm}{0.4cm} 1+ f^{\prime\prime}\right) + f^\prime \left(\rule{0mm}{0.4cm} D^i Q_j
\right.\right.
\nonumber \\ 
&& 
\!\!\!\!\!\!\!\!\!\!\!\!\!\!
\left.\left.
+ \Pi K^{i}_{\,\,\,j} \rule{0mm}{0.4cm}\right) \rule{0mm}{0.6cm}\right]
- \frac{\delta_{\,\,\,j}^i \alpha}{2f\left(1 + \frac{3{f^\prime}^2}{2f}\right)}
\left(\rule{0mm}{0.6cm} Q^2-\Pi^2\right)\left( \frac{{f^\prime}^2}{2f}
- f^{\prime\prime}\right) 
\nonumber \\
&& 
\!\!\!\!\!\!\!\!\!\!\!\!\!\!
= -\frac{\alpha}{2f\left(1 + \frac{3{f^\prime}^2}{2f}\right)}
\left\{ 2 S_{{\rm matt}\,\,\,j}^i \left(1 + \frac{3{f^\prime}^2}{2f} \right)
+ \delta_{\,\,\,j}^i \left[\rule{0mm}{0.6cm} f^\prime V^\prime
\right. \right.
\nonumber \\
&& 
\!\!\!\!\!\!\!\!\!\!\!\!\!\!
\left.\left. 
+ 2V\left(1 + \frac{{f^\prime}^2}{2f} 
\right) - \left(\rule{0mm}{0.4cm} S_{\rm matt}
- E_{\rm matt}\right) \left(1 + \frac{{f^\prime}^2}{f} \right)
\right]\right\}\,
\end{eqnarray}
where $S^{ab }_{\rm matt}:=\,\gamma_{\,\,\,c}^a \gamma_{\,\,\,d}^b
T^{cd}_{\rm matt}$ and $S_{\rm matt}$ is its trace.

Finally, equation~(\ref{KG}) can be written as the following first
order evolution equation~\cite{Salgado06}:
\begin{eqnarray}
\label{evPi1}
&&
\!\!\!\!\!\!\!\!\!\!\!\!
{\cal L}_{\mbox{\boldmath{$n$}}} \Pi  - \Pi K - Q^c D_c[{\rm ln}\alpha] 
- D_c Q^c \nonumber \\ 
&&
\!\!\!\!\!\!\!\!\!\!\!\!
= - \frac{ f V^\prime - 2f^\prime V -\frac{1}{2}f^\prime
\left( 1 +  3f^{\prime\prime} 
\right)\left(\rule{0mm}{0.4cm} Q^2-\Pi^2\right)
+ \frac{1}{2}f^\prime T_{{\rm matt}} }
{f\left(1 + \frac{3{f^\prime}^2}{2f}\right) } \,\,\,,\nonumber \\
\end{eqnarray}
where $T_{{\rm matt}}= S_{{\rm matt}}-E_{{\rm matt}}$. Here one must
remember that indices of 3-tensors (including 3-vectors) are raised
and lowered with the 3-metric $\mbox{\boldmath{$\gamma$}}$, and that
the contravariant time components of such quantities are identically
null. This is why one is usually only interested in the spatial
components of 3-tensors.

A useful evolution equation for the trace of the extrinsic curvature
takes the form:
\begin{eqnarray}  
\label{EDKSTT}
& & \!\!\!\!\!\!\!
 \partial_t K -\beta^l \partial_l K + \,^3\Delta \alpha
- \alpha K_{ij} K^{ij} 
- \frac{\alpha f^\prime}{f} \left(\rule{0mm}{0.4cm} D_l Q^l
+ \Pi K \right) \nonumber \\
& & \!\!\!\!\!\!\!
-\frac{\alpha}{f\left(1 + \frac{3{f^\prime}^2}{2f}\right)}
\left\{ \Pi^2 \left( 1 + \frac{3{f^\prime}^2}{4f}
+ \frac{3f^{\prime\prime}}{2} \right) \right. \nonumber \\
& & \!\!\!\!\!\!\! 
\left. + Q^2\left[ \frac{3{f^\prime}^2}{4f}\left(\rule{0mm}{0.4cm} 1
+ 2 f^{\prime\prime}\right)  
- \frac{f^{\prime\prime}}{2}\right]  \right\} \nonumber \\
& & \!\!\!\!\!\!\!
= \frac{\alpha}{2f\left(1 + \frac{3{f^\prime}^2}{2f}\right)}
\left\{ S_{\rm matt} + E_{\rm matt} \left(1 + \frac{3{f^\prime}^2}{f}\right) \right.\nonumber \\
& & \!\!\!\!\!\!\!
\left.
- 2V\left(1- \frac{3{f^\prime}^2}{2f}\right) - 3 f^\prime V^\prime \right\}
 \,\,\,,
\end{eqnarray}
where $\,^3\Delta:=D^l D_l$ is the Laplacian compatible with the
3-metric, and $D_c Q^c = \,\!\partial_c Q^c + Q^c \partial_c ({\rm
ln}\sqrt{\gamma})$ (here $\gamma={\rm det}\gamma_{ij}$).

The system of equations~(\ref{Pidef}), (\ref{EvQ}) and~(\ref{K_ab})
can now be rewritten as
\begin{eqnarray}  
\partial_t \phi &=& \alpha\Pi+\beta^a Q_a  \,\,\,\label{Pi}\,\,\,,\\
\label{evQ2}
\partial_t Q_i &-& \beta^l\partial_l Q_i - Q_l\partial_i \beta^l = D_i (\alpha\Pi)  \,\,\,,\\
\label{K_ij}
\partial_t \gamma_{ij} &=& -2\left( \alpha K_{ij} - \,D_{(i} \beta_{j)} \right) \,\,\,.
\end{eqnarray}
This system of evolution equations is to be completed with appropriate
evolution equation for the gauge variables (lapse and shift). This
issue is considered next.
\bigskip

In order to obtain a closed evolution system one has to impose gauge
conditions for the time variable $t$ and for the spatial coordinates
$x^i$. In this work we shall consider a modified Bona-Mass\'o time
slicing defined in the following way (cf. Ref.~\cite{Salgado06})
\begin{equation}
\label{STTBM}
\Box t = \left(\frac{1}{f_{BM}}-1\right)\,n^b n^c \nabla_b \nabla_c t 
+ \frac{\Theta}{f_{BM}} \: \frac{f^{\prime}}{f} \: \Pi\, n^a \nabla_a t \,,
\end{equation}
with $f_{BM}=f_{BM}(\alpha)>0$ a positive but otherwise arbitrary
function of the lapse, and $\Theta=\Theta(f_{BM})$ is in principle an arbitrary 
function of $f_{BM}$ which can {\em a posteriori}\/ be fixed so as 
to ensure a well behaved hyperbolic system (see
Section~\ref{sec:hyper} below). It is important not to confuse the
gauge function $f_{BM}(\alpha)$ with the NMC $f(\phi)$. 

Concerning the spatial coordinates, we shall consider the shift vector
as an {\em a priori}\/ known function of the coordinates.  However, in
the future it would be interesting to investigate some ``live'' shift
conditions and their effects in phenomena involving STT.

Now, the slicing condition~(\ref{STTBM}) for local coordinates
$x^a=(t,x^i)$ adapted to the 3+1 foliation of the spacetime reduces
to
\begin{equation} 
\label{STTBMlapse}
\frac{d\alpha}{dt} = - \alpha^2 f_{BM} \left(\rule{0mm}{0.4cm}
K- \frac{\Theta}{f_{BM}} \: \frac{f^{\prime}}{f} \: \Pi\right) \,\,\,\,,
\end{equation}
where $d/dt:= \partial_t - {\cal L}_{\mbox{\boldmath{$\beta$}}}$.  The
specific choices $\Theta=f_{BM}\equiv 1$ correspond to a modified harmonic
slicing condition (termed ``pseudoharmonic'' in~\cite{Salgado06}), which was
specially useful for the second order hyperbolicity analysis performed
in~\cite{Salgado06}. On the other hand, with $\Theta\equiv 0$ one recovers
the usual BM slicing condition. Just like in standard GR, we shall see
that one can take in general any positive $f_{BM}$ (not necessarily
$f_{BM}\equiv 1$) to obtain a well behaved strongly hyperbolic
system. Moreover, the choice $\Theta=1$ or $\Theta=f_{BM}$ provides 
simple sufficient conditions for regular eigenfields in the STT, as we will
show below. An inadequate choice for $\Theta$ leads to eigenfields that can
be ill-defined when $f_{BM}=1$ locally or globally (cf. Eqs.~(\ref{w3334}) and
(\ref{w2324}) below). A bad choice can be precisely $\Theta=0$, so 
the usual BM slicing condition seems not to be well adapted for
constructing strongly hyperbolic formulations of STT's in the Jordan 
frame.


\section{Hyperbolic systems}
\label{sec:hyper}

In the previous Section we showed a constrained system of first order
in time partial differential equations (PDE) corresponding to the
STT. This system is well formulated in the sense that one can
construct an unambiguous numerical algorithm that evolves initial data
satisfying the constraint equations. In the case of a globally null
scalar field, the above system of equations reduces to the usual ADMY
system of equations of GR. Even if such a system is well formulated in
the above sense, it is well-known that the evolution is not stable in
many cases.

It took several years to realize that such instabilities were not
associated with the numerical algorithms but rather with the
mathematical structure of the ADMY system itself. During the mid 90's,
many researchers started to suspect that the numerical instabilities
that seemed to plague several testbed codes based in the usual 3+1
formalism were caused by the lack of (strong) hyperbolicity of the
system. Therefore, many investigations were devoted to develop new
strongly hyperbolic formulations of the 3+1 evolution equations of GR
(see~\cite{Baumgarte:1998te} and references therein). However, at that
time, it was still not totally clear that such new hyperbolic
formulations presented more computational advantages than the usual
ADMY system.  The breakthrough towards a real improvement of the
numerical stability was the use of a novel formulation of GR which
used a conformal formalism, today known as the BSSN
formulation~\cite{Baumgarte:1998te,Shibata95}. In~\cite{Baumgarte:1998te}
it was recognized that the new conformal system presented some
hyperbolic features not present in the original ADMY system, and
moreover it was stressed that the use of suitably defined connection
coefficients and the subsequent addition of the momentum constraints
(in order to eliminate some divergence operators) proved to be
essential for numerical stability. Nonetheless, it was after a more
systematic analysis of the BSSN system that a much better
understanding of its hyperbolicity properties was
obtained~\cite{Sarbach:2002bt}.  In connection with the hyperbolicity
properties of the 3+1 system, one needs to mention the gauge choice
({\em i.e.}, the prescription of the evolution of the gauge
variables).  In the formal developments and proofs of theorems about
the well-posedness of the Cauchy problem of GR, it was the use of the
harmonic gauge that allowed one to write the Einstein field equations
as a second order diagonal hyperbolic system of quasilinear
PDE's. Nevertheless, in numerical implementations, the use of the
(pure) harmonic gauge has proven not to be very successful, specially
in the strong gravity regime where one requires to avoid (or freeze)
the formation of singularities in the numerical grid. A big step
towards solving this problem was the implementation of new gauge
conditions ({\em e.g.} the so-called Bona-Mass\'o time slicing) in
hyperbolic formulations~\cite{Bona94b}. Currently, many hyperbolic
formulations, together with Bona-Mass\'o like slicing conditions, have
been proposed in the literature.

In this and the following sections we will propose two systems for the
STT based on the the 3+1 system of Sec.~\ref{sec:ADM} but written in
fully first order form along the lines of the BSSN and the BMSS
formulations, and show that they are indeed strongly hyperbolic. Our
aim for constructing two strongly hyperbolic formulations of the STT is to
understand the way in which the modified Bona-Mass\'o
condition~(\ref{STTBMlapse}) enters into both formulations, and how
robust is this new class of slicing conditions. In fact we will see
that both formulations are compatible with the choices $\Theta=1,f_{BM}$ in
that the eigenfields turn out to be well behaved without imposing any
further stringent conditions on the gauge function $f_{BM}$ (apart
from it being positive definite).

Let us first outline the general approach we intend to follow for both
formulations.  From the numerical point of view, it seems to be more
useful to recast a second order system of PDE of hyperbolic type into 
a full first order form as follows
\begin{equation}
\label{PDE}
\partial_t {\vec u} + \mathbb{M}^i \partial_i {\vec u}
= {\vec S}({\vec u}) \,\,\,,
\end{equation}
where ${\vec u}$ represents collectively the fundamental variables
(like the $\gamma_{ij}$'s, $K_{ij}$'s, etc.), $\mathbb{M}^i$ are called
the {\em characteristic}\/ matrices of the system, and ${\vec S}({\vec
u})$ represents {\em source}\/ terms which include only the
fundamental variables but not their derivatives. This system of PDE is
said to be {\em quasilinear}\/ since it is linear in the derivatives
but in general is non-linear in the ${\vec u}$.

A quasilinear system of PDE of the form~(\ref{PDE}) possesses a
well-posed Cauchy problem (in the Hadamard sense) only in very special
circumstances (see Refs.~\cite{Reula04,Reula:1998ty} for a review),
which depend on the structure of the characteristic matrices 
$\mathbb{M}^i$. In general, when one deals with such a system one can
choose a specific spatial direction of propagation of fields along
which one analyzes the hyperbolic character of the system.  For
instance, if the direction of propagation is specified by a unit
covector $v_i$, then the hyperbolic nature of the system will be
determined by the matrix $\mathbb{C}:= v_i \mathbb{M}^i$, known as the
{\em principal symbol}. Namely, the system will be weakly hyperbolic
or strongly hyperbolic, if the following properties are satisfied
respectively: 1) $\mathbb{C}$ has real eigenvalues but an incomplete
set of eigenvectors; 2) $\mathbb{C}$ has real eigenvalues and a
complete set of eigenvectors (for any $v_i$). In addition, the system
is said to be symmetric or symmetrizable hyperbolic, respectively, if:
3) $\mathbb{M}^i$ are symmetric, 4) $\mathbb{M}^i$ can be
symmetrized. Only strongly, symmetric and symmetrizable hyperbolic
systems of PDE admit a well-posed Cauchy problem.

The systematic approach to analyze the hyperbolicity of a system of
evolution equations along these lines consists then in first recasting
the system in the form~(\ref{PDE}), and then solving the eigenvalue
problem of the corresponding principal symbol. If $\mathbb{C}$
satisfies the conditions `2)' above then one can ensure
well-posedness. In general, one finds that for the systems of
evolution equations found in the 3+1 formulation of GR, it is
necessary to add multiples of the constraint equations~(\ref{CEHfSST})
and~(\ref{CEMfSST}) to some of the evolution equations in order to
make the system strongly hyperbolic~\cite{Kidder01a}. This procedure
does not affect the physical solutions ({\em i.e.} those that satisfy
the constraint equations), but it does affect the structure of the
$\mathbb{C}$ matrix.

Now, suppose that the following system is strongly hyperbolic in a
given direction $v_i$ ({\em i.e.}, in what follows we will assume that
$\mathbb{C}$ satisfies the condition `2)' above).  From Eq.~(\ref{PDE})
one finds that
\begin{equation}
\label{PDE2}
\partial_t {\vec u} + \mathbb{C} \nabla_v {\vec u} \simeq 0\,\,\,,
\end{equation}
where $\nabla_v$ stands for the directional derivative along $v_i$,
and $\simeq$ indicates that we shall focus only in the principal part
of (\ref{PDE}).  If we define now $\mathbb{R}$ as the matrix of
eigenvectors of $\mathbb{C}$, then $\mathbb{L} = \mathbb{R}^{-1}
\mathbb{C} \mathbb{R}$ is the diagonal eigenvalue matrix. Notice that
$\mathbb{R}$ can clearly be inverted since we have assumed that
$\mathbb{C}$ has a complete set of eigenvectors. Multiplication of
(\ref{PDE2}) on the left by $\mathbb{R}^{-1}$ leads to the equivalent
system
\begin{equation}
\label{PDE3}
\partial_t {\vec w} + \mathbb{L} \nabla_s {\vec w} \simeq 0\,\,\,,
\end{equation}
where we have introduced the functions ${\vec w}:= \mathbb{R}^{-1}
{\vec u}$ which are called the {\em eigenfields} or {\em
eigenfunctions}.  Clearly the new system (\ref{PDE3}) is now decoupled
and each eigenfield $w_i$ propagates with its own characteristic speed
given by its corresponding eigenvalue $\lambda_i$.

All this construction depends heavily on the existence of a complete
set of eigenvectors (in addition to the existence of real
eigenvalues). The system (\ref{PDE2}) is therefore equivalent to
(\ref{PDE3}) when the strongly hyperbolicity conditions are
met. Furthermore, conditions on the smoothness of the eigenvectors (or
eigenfields) and several other conditions concerning the gauge speeds
({\em e.g.}, real-valued speeds, singularity avoidance) impose in turn
conditions on the class of gauges adopted in the selection of the time
slicing (see below) and the spatial coordinates (although here we
shall consider the shift condition as a prescribed one).

Now, it turns out that very often (depending on the complexity of the
system) one can construct the equivalent system~(\ref{PDE3}) not by
using direct methods to diagonalize the matrices $\mathbb{M}^i$, but
instead by a ``judicious guessing'' approach which consists in
constructing the eigenfields and corresponding eigenvalues by
inspection (see Appendix B for a simple example). This is for instance
the approach followed in~\cite{Bona94b}.  In sections IIIA and IIIB
below we use this inspection approach to construct the eigenfields and
their corresponding speeds of propagation.

\subsection{Bona-Mass\'o Formulation of STT}

In order to construct a full first order system based on the system of
equations of Sec.II, we first need to define new first order variables
in the following way
\begin{equation}
\label{ai,dkij}
a_i=\partial_i\,{\rm ln} \alpha \,\,\,,
\hspace{0.8cm}
d_{kij}=\frac{1}{2}\,\partial_k\,\gamma_{ij}\,\,\,.
\end{equation}
Moreover, we will also introduce the auxiliary variable
\begin{equation}
V_i = {d_{im}}^{m} - {d^{m}}_{mi} \,\,\,,
\end{equation}
where ${d_{ij}}^{l}:= \gamma^{lm}d_{ijm}$ and ${d^{l}}_{ij} :=
\gamma^{lm}d_{mij}$. Using this definition, the principal part of
$\,^3 R_{ij}$ reads
\begin{equation}
\,^3 R_{ij} \simeq - \partial_m\, {d^{m}}_{ij} - \partial_{(i}
\, \left( \rule{0mm}{0.4cm} 2 V_{j)} - {d_{j)\,m}}^{m} \right) \,.
\end{equation}
With such definitions the evolution equations~(\ref{EDEfSST}) will be
of first order in space for $a_i$ and $d_{kij}$, instead of second
order for the lapse and the 3-metric.

After some algebra one obtains the following set of 34 first order
evolution equations, written here only up to principal part (the full
form of the equations can be seen in the Appendix A):
\begin{eqnarray}
\partial_0 Q_i  &\simeq& \alpha \: \partial_l \left( \delta^l_i \: \Pi \right)
\,, \label{eq:evQ}\\
\partial_0 \Pi &\simeq& \alpha \: \partial_l Q^l \,, \label{eq:evPi} \\
\partial_0 a_i &\simeq&-\alpha \:\partial_l \left[ \delta^l_i
\left(f_{BM} \: K - \Theta \: \frac{f^\prime}{f}\,\Pi \right) \right] \,, \\ 
\partial_0 {d}_{ijk} &\simeq& - \alpha \: \partial_i K_{jk} \,, \\
\partial_0 K_{ij} &\simeq& - \alpha \: \partial_l {\Lambda^l}_{ij} \,, \\
\label{eq:evKij}
\partial_0 V^{i} &\simeq& \alpha \: \partial_l \left[
\left( 1 + \varsigma \right) \left( K^{il} - \gamma^{il} K \right)
+ \varsigma \gamma^{il} \frac{f^\prime}{f} \Pi \right] \,, \hspace{10mm}
\label{eq:evVi}
\end{eqnarray}
where $\partial_0:= \partial_t - \beta^l\partial_l$, $V^i:=
\gamma^{il}V_l$ and
\begin{equation}
{{\Lambda}^l}_{ij} := {d^{l}}_{ij} + \delta^l_{(i} \left( {a_{j)}}
+ \frac{f^\prime}{f} \: Q_{j)} + 2 V_{j)} - {d_{j)\,m}}^{m} \right)
\label{eq_Lambda} \,.
\end{equation}
Though the ${{\Lambda}^l}_{ij}$ are not independent variables, it is
nevertheless useful to have their evolution equation which take the
form
\begin{eqnarray}
\label{evlambda}
&& \!\!\!\!\!
 \partial_0\,{{\Lambda}^l}_{ij}\,\simeq  -\alpha \gamma^{lm}\partial_m K_{ij} + 
\alpha \delta^l_{(i} \partial_m \left[\rule{0mm}{0.5cm} 2(1+\varsigma)\left(K^{m}_{\,\,\,\,j)} \right.\right.
\nonumber \\
&& \!\!\!\!\!
\left.\left. - \delta^{m}_{\,\,\,\,j)} K\right) +
\delta^{m}_{\,\,\,\,j)}(1-f_{BM}) K + \frac{f^\prime}{f}\,\Pi \delta^{m}_{\,\,\,\,j)} (1+\Theta + 2\varsigma)\right],
\nonumber \\
\end{eqnarray}
Here we must stress the fact that in Eq.~(\ref{eq:evVi}) we have added
a multiple $\alpha \varsigma$ of the momentum constraint
(\ref{CEMfSST}). Moreover, the evolution equations for the lapse and
the 3-metric were not included since they are ``trivial'' in the sense
that the lapse and the 3-metric do not evolve up to principal part and
only they propagate along the ``normal lines''. As pointed out
in~\cite{Bona97a}, the above evolution system,
Eqs.~(\ref{eq:evQ})$-$(\ref{eq:evVi}), can be seen as a reduced system
evolving in a inhomogeneous ``background''.

Now, in order to obtain the eigenfields propagating in one specific
direction we will consider the `$x$' direction, {\em i.e.}
$v_i=(1,0,0)$, and ignore the derivatives in the other directions
(cf.~\cite{Alcubierre08a}).  This is equivalent to analyzing only the
characteristic matrix $\mathbb{M}^x$. Thus, for $q \neq x$, it is clear
from the evolution equations~(\ref{eq:evQ})$-$(\ref{eq:evVi}) that
there are 18 eigenfunctions which propagate with speed $-\beta^x$,
namely $w_{1,2}=Q_q$ (corresponding to the two eigenfields propagating
in the two directions orthogonal to `$x$'), $w_{3,4}=a_q$,
$w_{5-16}={d}_{qjk}$ (12 eigenfields: six for each $q$), and
$w_{17,18}=V_q + (1+\varsigma)d^{l}_{\,\,\,lq}$ .

Furthermore, by using the evolution equations we appreciate that the
following combinations
\begin{eqnarray}
w_{19}= &&a_x-\,f_{BM}\,{d_{xm}}^{m}-\,\Theta\,\frac{f^\prime}{f}\,Q_x\,\,, \\
\label{w20}
w_{20}= &&V^x -\varsigma \frac{f^\prime}{f}\,Q^x
- \left( 1+\varsigma\right)\hat \Lambda^{x} \,\,,
\end{eqnarray}
provide two more eigenfunctions propagating along the normal lines with
$-\beta^x$ speed, where
\begin{equation}
\label{hatLambda}
\hat \Lambda^{x}:= \Lambda^{x}-\frac{\Lambda^{xxx}}{\gamma^{xx}}
=\left(\rule{0mm}{0.4cm}\gamma^{xx}\gamma^{pq}  -\gamma^{xp}\gamma^{xq}\right) \frac{\Lambda^{x}_{\;pq}}{\gamma^{xx}} \,\,\,,
\end{equation}
is the projection of $\Lambda^{x}_{\;pq}$ on the surface $x={\rm
const.}$ (for $q\neq x$ and $p\neq x$) \cite{Alcubierre08a}. Here we
used $\Lambda^{x}:= \gamma^{lm} {{\Lambda}^x}_{lm}$ and
$\Lambda^{xxx}:= \gamma^{xl}\gamma^{xm}{{\Lambda}^x}_{lm}$. Such a
projection is in fact achieved with $(P^x)_{ij}:= \gamma_{ij}- (s^x)_i
(s^x)_j$, where ${\vec s}^{\;x}$ is the normal to the surface with
components given by $(s^x)_i= \delta^x_i/\sqrt{\gamma^{xx}}$ and
$(s^x)^i= \gamma^{xi}/\sqrt{\gamma^{xx}}$. So $\hat \Lambda^{x} =
(P^x)^{ij} {{\Lambda}^x}_{ij}$. In fact by projecting
Eq.~(\ref{evlambda}) on the surface $x={\rm const.}$, and focusing in
propagation only in the `$x$' direction, one obtains $\partial_0 \hat
\Lambda^{x}\simeq -\alpha\gamma^{xx} \partial_x \hat K$, where $\hat
K:= (P^x)^{ij} K_{ij}$ whose expression is given below in
Eq.~(\ref{hatK}).

Raising the index in Eq.~(\ref{eq:evQ}), and combining the evolution
equation for $Q^x$ with Eq.~(\ref{eq:evPi}), we obtain the following
two eigenfunctions
\begin{equation}
w_{21,22}= Q^x\pm\,\sqrt{\gamma^{xx}}\,\Pi \,\,\,,
\end{equation}
which propagate along the light cones with eigenvalues 
\begin{equation}
\label{l2122}
\lambda_\pm= -\beta^x \mp\,\alpha\,\sqrt{\gamma^{xx}} \,\,\,.
\end{equation}
Next, combining the equations for $K_{qp}$ and ${\Lambda^x}_{qp}$, we
form the following six more eigenfunctions:
\begin{eqnarray}
w_{23-28}= && {\Lambda^x}_{pq} \pm\,\sqrt{\gamma^{xx}}\,K_{pq} \,\,\,,
\end{eqnarray}
which propagate also along the light cones with eigenvalues
\begin{equation}
\label{l23-28}
\sigma_\pm= -\beta^x \pm\,\alpha\,\sqrt{{\gamma}^{xx}} \,\,\,.
\end{equation}
Combining the expressions for ${K^x}_q$ and ${\Lambda^{xx}}_{q}$
(${\Lambda^{xx}}_{q}:= \gamma^{xl}{{\Lambda}^x}_{lq}$), one can see
that
\begin{equation}
\label{w2932}
w_{29-32}={\Lambda^{xx}}_{q} \pm \sqrt{-\varsigma \,\gamma^{xx}}\,{K^x}_q \,\,\,,
\end{equation}
are four more eigenfunctions with eigenvalues
\begin{equation}
\label{l29-32}
\omega_\pm= -\beta^x \pm\alpha\,\sqrt{-\varsigma \, \gamma^{xx}} \,\,\,.
\end{equation}
Finally the last 2 eigenfunctions turn out to be
\begin{eqnarray}
\label{w3334}
&&
w_{33,34} = \Lambda^x + \frac{2(1+\varsigma)}{f_{BM}-1} \hat \Lambda^x \nonumber \\
&&
\pm  \sqrt{f_{BM}\,\gamma^{xx}} \left[ K + \frac{2(1+\varsigma)}{f_{BM}-1} \hat K\right] \nonumber \\
&&
+ \left(\frac{2\varsigma +1 + \Theta}{f_{BM}-1}\right) \frac{f^\prime}{f}\,\left( Q^{x} \mp \sqrt{f_{BM} \gamma^{xx}} \Pi\right) \, , \quad
\end{eqnarray}
where 
\begin{equation}
\label{hatK}
\hat K:= K-\frac{K^{xx}}{\gamma^{xx}}
=\left(\rule{0mm}{0.4cm}\gamma^{xx}\gamma^{pq}  -\gamma^{xp}\gamma^{xq}\right) \frac{K_{pq}}{\gamma^{xx}} \,\,\,,
\end{equation}
is defined in a similar fashion as $\hat \Lambda^x$
(cf. Eq.~(\ref{hatLambda})).

The eigenfields $w_{33,34}$ propagate with the ``gauge speeds'' given
by the following eigenvalues
\begin{equation}
\label{l3334}
\delta_\pm= -\beta^x \pm\alpha\,\sqrt{f_{BM}\gamma^{xx}} \,\,\,.
\end{equation}
We have then 34 linearly independent eigenfunctions $w_{1-34}$ which
are equivalent to the 34 linearly independent variables
($Q_i,\Pi,a_i,d_{ijk},K_{ij},V^i$), so the system is strongly
hyperbolic.

From the 34 eigenfunctions obtained above, we appreciate that only
$w_{33,34}$ might develop gauge divergences, while the remaining
eigenfields are smooth (as long as the original variables themselves
are smooth) regardless of the time slicing.  Note, for instance, how
in Eq.~(\ref{w3334}) the choice $f_{BM}\equiv 1$ can
cause those eigenfunctions to blowup unless we choose the value 
$\varsigma=-1$ (see the coefficients of $\hat \Lambda^x$ and $\hat K$ in
$w_{33,34}$). Moreover, even with $\varsigma=-1$, 
another blowup of the same sort might appear in the coefficient with $\frac{f^\prime}{f}$
of $w_{33,34}$. However, that problem can in turn be avoided by a suitable choice for 
$\Theta$. The simplest choices are $\Theta=1$ and $\Theta=f_{BM}$. For instance, 
the eigenfunctions $w_{33,34}$ with $\varsigma=-1$ and $\Theta=1$ simply become
\begin{equation}
\label{w3334sim}
w_{33,34}= \Lambda^{x}\pm  \sqrt{f_{BM}\,\gamma^{xx}} K \,\,\,.
\end{equation}
Moreover, with $\varsigma=-1$ the eigenfield (\ref{w20}) becomes
\begin{equation}
\label{w20p}
 w_{20}= V^x + \frac{f^\prime}{f}\,Q^x \,\,\,.
\end{equation}
The value $\varsigma=-1$ also turns out to be compatible with the
condition for real-valued eigenfunctions $w_{29-32}$
(cf. Eq.~(\ref{w2932})). Now, when one chooses $f_{BM}\equiv 1$, the two optimal 
values for $\Theta$ reduce both to $\Theta=1$, and the slicing 
condition~(\ref{STTBMlapse}) is compatible with the modified harmonic 
time slicing which together with the modified harmonic spatial coordinates 
(both written as $\Box x^a= -\frac{1}{f}\,\nabla^a f$) were the ones required 
in Ref. \cite{Salgado06} to reduce the field equations of STT (without matter) 
into a quasilinear diagonal second order hyperbolic system.
Furthermore, note that $w_{33,34}$ degenerate ($w_{33}=w_{34}$) if
$f_{BM}=0$, and the eigenfields are not complete in that case. We then
appreciate the importance of the condition $f_{BM}>0$. In particular,
in the BMSS formulation of standard GR, the condition $f_{BM}\geq 1$ 
seemed to be the most adequate for a convenient mode propagation and
singularity avoidance behavior.

Finally, let us remark that in pure GR the eigenfield (\ref{w20p}) is
given by $w_{20}= V^x$ which propagates along the normal lines. We just
saw that $w_{17,18}= V_q$ (with $\varsigma=-1$) also propagate along
the normal lines.  Therefore in pure GR all three $V_i$'s propagate
along the normal lines. In pure GR there are also only $30$ eigenfields
($w_{1,2}$ and $w_{21,22}$ are absent), and in this limit all the
eigenfields coincide or are equivalent to the ones obtained in
Ref.~\cite{Alcubierre08a}, which are precisely the eigenfields found
in the Bona-Mass\'o formulation. As emphasized in
Ref.~\cite{Alcubierre08a}, in GR taking $\varsigma=0$ as opposed to
$\varsigma=-1$ (which is the value proposed in the Bona-Mass\'o
formulation), is equivalent to the full first order ADMY system. That
is, $\varsigma=0$ amounts to not adding the momentum constraints to
the evolution equations. The ADMY system is then weakly hyperbolic.
This feature is manifest from Eq.~(\ref{w2932}) in that one has two
less independent eigenfields with $\varsigma=0$, so that one cannot
reconstruct the original 30 variables ($a_i,d_{ijk},K_{ij},V^i$) of
GR.  We then conclude that for STT $\varsigma=0$ also leads to a
weakly hyperbolic system since then one has only 32 eigenfunctions
from which it is impossible to reconstruct the 34 field variables
listed above.


\subsection{BSSN Formulation of STT}

The BSSN formulation of GR~\cite{Baumgarte:1998te,Shibata95} is
perhaps the most popular and useful formulation for numerical
approaches in use today. This formulation was found to be much more
stable in numerical simulations than the standard ADMY approach. A
further analysis of the BSSN formulation revealed that this is in fact
a strongly hyperbolic system~\cite{Sarbach:2002bt}. Although it is
clear that the strong hyperbolicity property is a key ingredient that
causes BSSN to perform better than the ADMY formulation (which is only
weakly hyperbolic), this feature seems to be not the only issue in the
assessment. In fact there are other strongly hyperbolic formulations
that perform poorly in comparison with BSSN. Apparently it is the
conformal approach that provides another important ingredient to the
numerical stability. Nevertheless, it is still not very well
understood why the conformal approach makes such an improvement
relative to other strongly hyperbolic formulations.  More recently the
BSSN formulation has also shown great stability in binary black hole
simulations when the moving puncture method is
implemented~\cite{Campanelli:2005dd,Baker:2005vv}. Notably, this
formulation is able to evolve the initial data of two inspiraling
black holes until the merger stage and beyond without the code
crashing. This was a long standing problem in numerical relativity.
Moreover, this formulation has also been very successful when matter
is included (see for instance~\cite{Tung-Liu07,Baiotti07,Shibata06a}).

In view of these advantages, we propose here an analog of the BSSN
formulation but extended to the STT. Moreover, we will also deal with
the proposed modified Bona-Mass\'o slicing condition in this formulation and study the
roll played by the function $\Theta$ in the construction of well
behaved eigenfields. Just like in Sec. IIIA, we shall define new
variables so as to obtain a full first order system (both in time and
space), but following now the conformal approach of BSSN.

We consider again the first of Eqs.~(\ref{ai,dkij}), and introduce the
following variables:
\begin{eqnarray}
\label{tildegamma}
\tilde{\gamma}_{ij} &=& e^{-4\psi}\,\gamma_{ij}\,\,,\\
\label{tildeAij}
\tilde{A}_{ij} &=& e^{-\zeta\,\psi}\,A_{ij} \,\,,\\
\label{tilded}
\tilde{d}_{ijk} &=& \frac{1}{2} \partial_i\,\tilde{\gamma}_{jk} \,\,,\\
\Psi_i &=& \partial_i\,\psi \,\,,\\
\label{tilde3Gamma}
^3\tilde{\Gamma}^{i} &=& \tilde{\gamma}^{jk}\,^3\tilde{\Gamma}^{i}_{jk} =  -\partial_j\,\tilde{\gamma}^{ij}\,\,,
\end{eqnarray}
where $\psi$ is chosen such that the determinant $\tilde{\gamma}$ of
the conformal metric is equal to one~\footnote{In fact, as suggested
in Ref.~\cite{Gourgoulhon07}, when one works with coordinates other
that Cartesian-type, it turns out to be better to define $\psi$ so
that ${\rm det} \tilde \gamma_{ij}= {\rm det} \tilde f_{ij}$, where
$f_{ij}$ represents a background Riemannian metric.  For instance if
$\Sigma_t$ is asymptotically flat one can take $f_{ij}$ to be the flat
metric in some coordinates adapted to the symmetry of the problem. In
spherical symmetry, say, \mbox{$f_{ij}= {\rm diag} (1,r^2, r^2
\sin^2\theta)$} and ${\rm det} \tilde \gamma_{ij}=r^4 \sin^2
\theta$.  This approach has the advantage that $\psi$ becomes a true
scalar while all the tensorial quantities defined using $\tilde
\gamma_{ij}$ and $\psi$ become true tensors instead of tensor
densities.}, $^3\tilde{\Gamma}^{i}_{jk}$ are the 3-Christoffel symbols
associated with the conformal metric, $\zeta$ is a constant, and
$A_{ij}$ is the trace-free part of $K_{ij}$ given by:
\begin{equation}
\label{Aij}
{A}_{ij}=K_{ij} - \frac{1}{3}\,\gamma_{ij}\,K  \,\,.
\end{equation}
From the evolution system of Sec.II one obtains the following
conformal evolution system up to principal part (see the Appendix A
for the full system of equations):
\begin{eqnarray}
&\partial_0 Q_i\,\simeq&\alpha\,\partial_l\,\left(\delta^l_i\,\Pi\right)\,\,\,, \\
&\partial_0 \Pi\,\simeq&\alpha\,\partial_l\,Q^l\,\,\,, \\
&\partial_0 a_i\,\simeq&-\alpha\,\partial_l\,\left[\delta^l_i\,\left(f_{BM}\,K-\Theta\,\frac{f^\prime}{f}
\,\Pi\right)\right]\,, \\ 
&\partial_0 \tilde{d}_{ijk}\,\simeq&-\alpha\, e^{\left( \zeta -4 \right) \psi}\,\partial_i\,\tilde{A}_{jk}\,\,\,, \\
&\partial_0 \Psi_i\,\simeq&-\frac16\,\alpha\,\partial_i\,K\,\,\,, \\
&\partial_0 K\,\simeq&-\alpha\,e^{-4\,\psi}\,\tilde{\gamma}^{jl}\partial_l\,\left(a_j-\frac{f^\prime}{f}\,Q_j\right)\,\,\,, \\
&\partial_0 \tilde{A}_{ij}\,\simeq&-\alpha\,\,e^{-\zeta\,\psi}\,\partial_l\,{\tilde{\Lambda}^l}_{\,\,ij}\,\,\,, \\
\nonumber \\
\label{Gamtilde}
&\partial_0\,\!^3\tilde{\Gamma}^{i}\,\simeq&\alpha\,\partial_l\,\left[\rule{0mm}{0.6cm}\left(\xi - 2\right)\,e^{\left( \zeta -4 \right) \psi}\, \tilde{A}^{il}  \right. \nonumber \\
&&
\left.
+ \tilde{\gamma}^{il}\,\xi\,\left(-\frac{2}{3}\, K + \frac{f^\prime}{f} \Pi\right) \rule{0mm}{0.6cm}\right]\,\,\, .
\end{eqnarray}
Again, we stress that in Eq.~(\ref{Gamtilde}) above we have added a
multiple $\alpha \xi$ of the momentum constraints~(\ref{CEMfSST}) to the
evolution equation for the variable $^3\tilde{\Gamma}^{i}$.  Notice
also $\tilde{A}^{ij}= \tilde{\gamma}^{il} \tilde{\gamma}^{jm}
\tilde{A}_{lm} =e^{\left( 8- \zeta\right) \psi} {A}^{ij}$, and we have
also defined
\begin{equation}
{\tilde{\Lambda}^l}_{\,\,ij}:=\left[{\tilde{d}^{l}}_{\,\,ij} + \delta^l_{(i}\,\left(\,{a_{j)}}+\frac{f^\prime}{f}\,Q_{j)}-\,^3\tilde{\Gamma}_{j)}+2\Psi_{j)} \right)\right]^{\rm TF}
\label{eq_Lambda1}\,\,\,,
\end{equation}
where $\rm{TF}$ means that the quantity between the brackets is
trace-free ({\em i.e.} $\tilde{\gamma}^{ij}
{\tilde{\Lambda}^l}_{\,\,ij}\equiv 0$). Here $\tilde{d}^{l}_{\,\,ij}
:= \tilde{\gamma}^{lm}\tilde{d}_{mij}$ and $\,^3\tilde{\Gamma}_{j}:=\tilde{\gamma}_{jl}\,^3\tilde{\Gamma}^{l}$. For completeness we provide the evolution equation for ${\tilde{\Lambda}^l}_{\,\,ij}$:
\begin{eqnarray}
&\!\!\!\!\!
 \partial_0\,{\tilde{\Lambda}^l}_{\,\,ij}\,\simeq-\alpha\,\left\{\rule{0mm}{0.6cm}
e^{\left( \zeta -4 \right) \psi}\partial_m\left[
\tilde{\gamma}^{lm}\tilde{A}_{ij} + \left(\xi-2\right)\,\delta^l_{(i}\,{\tilde{A}^m}_{\,\,\,j)} \right] \right.\nonumber \\
& \left. 
\!\!\!
+\left(f_{BM} +\frac{1-2\xi}{3}\right) \,\delta^l_{(i}\,\partial_{j)}K - \frac{f^\prime}{f}\left(\Theta+1-\xi\right)
\,\delta^l_{(i}\,\partial_{j)}\Pi \rule{0mm}{0.6cm}\right\}^{\rm TF} 
\!\!\!\!\!\!\nonumber \\
\end{eqnarray}

In the above system of evolution equations we consider only the
reduced system since the lapse and the 3-metric propagate along the
normal lines.

Like in the BM hyperbolic formulation of Sec.IIIA, there are again 34
variables, namely 15 linearly independent
$\tilde{d}_{ijm}$, five linearly independent $\tilde{A}_{ij}$
\footnote{We remind the reader that $\tilde A_{ij}$ and $\tilde{d}_{ijm}$ have
only 5 and 15 linearly independent components respectively, since both
are traceless. In fact, from Eq.~(\ref{tilded}) we conclude that
$\tilde{\gamma}^{jm} \tilde{d}_{ijm}=\frac{1}{2} \: \partial_i ({\rm
det} \tilde{\gamma}_{ij}) \equiv 0$, since one chooses
$\tilde{\gamma}_{ij}$ so that ${\rm det} \tilde{\gamma}_{ij}\equiv
1$.}, three $^3\tilde{\Gamma}^i$, one $K$, three $\Psi_i$ (for the
derivatives of the conformal factor), three $a_i$, and finally, four
variables $\Pi$ and $Q_i$ related to the derivatives of the scalar
field.

In order to find the eigenfields associated with these 34 variables,
we consider again a particular direction of propagation, say
$i=x$. For $q\neq x$, it is clear from the above evolution equations
that there are again 18 eigenfunctions which propagate with speed
$-\beta^x$, namely $w_{1,2}=Q_q$, $w_{3,4}=a_q$, $w_{5,6}=\Psi_q$, and
$w_{7-16}=\tilde{d}_{qjk}$.

Furthermore, we appreciate that the following combinations
\begin{eqnarray}
w_{17}=&& \!\!\!\!
a_x-6\,f_{BM}\,\Psi_x-\,\Theta\,\frac{f^\prime}{f}\,Q_x\,\,\,, \\
&& \nonumber \\
\label{w17i}
w_{18-20}=&&\,\!\!\!\!
\!^3\tilde{\Gamma}^i+\left(\xi-2\right)\,{{\tilde{d}_m}}^{\;\;\;mi} 
- \xi \,\tilde{\gamma}^{ik}\,\left(4\,\Psi_k + \frac{f^\prime}{f}\,Q_k\right)\,\,\,,\nonumber \\
\end{eqnarray}
provide another four eigenfunctions with speed $-\beta^x$, where
${{\tilde{d}_m}}^{\;\;\;mi}:=\tilde{\gamma}^{ij}\tilde{\gamma}^{ml}
\tilde d_{mlj}$. Thus, we have obtained 20 eigenfunctions.

Moreover, just like in Sec.IIIA, we obtain another two eigenfunctions
from the following combinations
\begin{equation}
w_{21,22}=Q^x\pm\,\sqrt{\gamma^{xx}}\,\Pi\,\,\,,
\end{equation}
with eigenvalues
\begin{equation}
\lambda_\pm= -\beta^x \mp\,\alpha\,\sqrt{\gamma^{xx}}\,\,\,.
\end{equation}
Next, combining the evolution equations for $K, Q^x, a^x$ and $\Pi$, we form the
following two eigenfunctions:
\begin{eqnarray}
&&
w_{23,24}= a^x -\left(\frac{\Theta-f_{BM} }{1-f_{BM}}\right)
\frac{f^\prime}{f}\,Q^x
\nonumber\\
\label{w2324}
&&
\pm\,e^{-2\,\psi} \sqrt{f_{BM}\,\tilde{\gamma}^{xx}}
\left[K + \left(\frac{\Theta-1}{1-f_{BM}}\right)\frac{f^\prime}{f}\,\Pi\right]\,\,\,
\end{eqnarray}
with eigenvalues 
\begin{equation}
\sigma_\pm= -\beta^x \pm\,\alpha\,e^{-2\,\psi}\,\sqrt{f_{BM}\,\tilde{\gamma}^{xx}}\,\,\,.
\end{equation}
Combining the evolution equations for
${\tilde\Lambda^{xx}}_{\,\,\,\,\,\,q}$
(${\tilde\Lambda^{li}}_{\,\,\,\,j}:=\tilde{\gamma}^{ik}{\tilde{\Lambda}^l}_{\,\,kj}$)
and ${\tilde{A}^x}_{\;\,q}$, one appreciates that
\begin{equation}
\label{w2528}
w_{25-28}={\tilde\Lambda^{xx}}_{\,\,\,\,\,\,q} \pm e^{\left( \zeta -2\right)\psi }\sqrt{\frac{\xi}{2}\,\tilde{\gamma}^{xx}}\,{\tilde{A}^x}_{\;q}\,\,\,,
\end{equation}
are four more eigenfunctions with eigenvalues
\begin{equation}
\omega_\pm= -\beta^x \pm\alpha\,e^{-2\psi } \sqrt{\frac{\xi}{2}\,\tilde{\gamma}^{xx}} \,\,\,.
\end{equation}
Furthermore, combining ${\tilde\Lambda^{xxx}}$
(${\tilde\Lambda^{ijk}}:=\tilde{\gamma}^{lk}{\tilde{\Lambda}}^{ij}_{\,\,\,\,l}$),
${\tilde\Lambda^{x}}_{\;qp}$, ${\tilde{A}^{xx}}$ and
${\tilde{A}}_{qp}$, we obtain the following eigenfunctions:
\begin{eqnarray}
\!\!\!\!\!\!\!\!\!
w_{29-32}&=& {\tilde\Lambda^{x}}_{\;pq}+\frac{{\tilde{\gamma}}_{pq}}{2{\tilde{\gamma}}^{xx}}{\tilde\Lambda^{xxx}} 
\nonumber\\ 
&&\pm e^{\left( \zeta -2\right)\psi }\sqrt{\tilde{\gamma}^{xx}}\,\left( {\tilde{A}}_{pq} + \frac{{\tilde{\gamma}}_{pq}}{2{\tilde{\gamma}}^{xx}} {\tilde{A}}^{xx}\right)\,,
\end{eqnarray}
with eigenvalues
\begin{equation}
\delta_\pm= -\beta^x \pm \alpha\,e^{-2\psi }\sqrt{\tilde{\gamma}^{xx}}\,\,\,.
\end{equation}
As remarked in~\cite{Alcubierre08a}, $w_{29-32}$ are in fact only four
independent eigenfunctions since these are symmetric and {\it surface
traceless}.

Finally the last 2 eigenfunctions are obtained by combining the
evolution equations for ${\tilde{\Lambda}^{xxx}}$, $a^x$, $Q^x$,
${\tilde{A}^{xx}}$, $K$ and $\Pi$:
\begin{eqnarray}
\label{w3334a}
&& \!\!\!\!\!\!\!\!\!
w_{33,34}= {\tilde\Lambda^{xxx}}\, -\, {\frac{1}{3}} {\tilde{\gamma}}^{xx} \left( 2 \tilde{a}^x +\frac{f^\prime}{f} \tilde{Q}^x \right) \nonumber \\
&& \!\!\!\!\!\!\!\!\!\!\!
\pm \, e^{2\psi}\sqrt{\tilde{\gamma}^{xx} \, {\frac{2\xi-1}{3}}}\,\left[\rule{0mm}{0.6cm} {e^{\left(\zeta-4 \right) \psi}\,\tilde{A}}^{xx} \!+\! {\tilde{\gamma}}^{xx}\left( {- \frac{2}{3}}  K + \frac{f^\prime}{f} \, \Pi  \, \right) \right]  ,
\nonumber \\  
\end{eqnarray}
where $\tilde{a}^x:= {\tilde{\gamma}}^{xl} a_l$, and $\tilde{Q}^x:=
{\tilde{\gamma}}^{xl} Q_l$.  These propagate with the following speeds
\begin{equation}
\label{eigenomega}
\eta_\pm= -\beta^x \pm \alpha\,e^{-2\psi}\sqrt{\tilde{\gamma}^{xx}{\frac{2\xi-1}{3}}}\,\,\,.
\end{equation}

Thus, we have found the complete set of eigenfunctions and
eigenvalues. The 34 linearly independent eigenfields $w_{1-34}$ are
equivalent to the 34 linearly independent variables
($Q_i,\Pi,a_i,\tilde d_{ijk},\Psi_i,K,\tilde A_{ij},\,\!^3\tilde
\Gamma^i$).

Now, like in the Bona-Mass\'o system of Sec.IIIA, there are some
eigenfunctions (namely $w_{23,24}$) that might blow-up when
$f_{BM}=1$. In order to avoid this problem while,
maintaining the possible choice $f_{BM}\equiv 1$, one has two simple possibilities: 
$\Theta=1,f_{BM}$. These are the same conditions found in Sec.IIIA for well behaved eigenfields. 
For instance, with $\Theta=1$ the eigenfunctions $w_{23,24}$ reduce to
\begin{equation}
w_{23,24}= a^x - \,\frac{f^\prime}{f}\,Q^x \pm\,e^{-2\,\psi} \sqrt{f_{BM}\,\tilde{\gamma}^{xx}}\,K \,\,\,,
\end{equation}
with the same eigenvalues, and the blow-up disappears even if 
$f_{BM}=1$. The same happens with the choice $\Theta=f_{BM}$ and thus 
we conclude again that in either case ($\Theta=1,f_{BM}$), the choice $f_{BM}\equiv 1$ requires 
$\Theta=1$ which lead, as emphasized before, to the modified 
harmonic slicing condition $\Box t= -\frac{1}{f}\,\partial^t f$ \cite{Salgado06} 
(cf. Eq.~(\ref{STTBM}) with $\Theta=f_{BM}\equiv 1$).

 The remaining eigenfunctions are smooth for any $f_{BM}>0$
and the eigenvalues are real valued. Note, on the other hand, that if
$f_{BM}=0$ the eigenfunctions $w_{23,24}$ degenerate and then the
system is no longer complete. 

None of the eigenvalues depend on $\zeta$, and the values $\zeta=2,4$
seem the optimal since in that case many exponential factors cancel
out ($\zeta=4$ is the one used in the original BSSN formulation of GR,
while $\zeta=-2$ has been used for constructing initial
data~\cite{Gourgoulhon07}). On the other hand, from
Eqs.~(\ref{w3334a}) and (\ref{eigenomega}) we appreciate that the eigenfunctions and the 
eigenvalues are real and nondegenerate only for $\xi>1/2$, and therefore the system is strongly hyperbolic if
this condition holds. A particular case corresponds to the value
$\xi=2$ which seems to be the most convenient since in that case the
eigenfunctions $w_{18-20}$ (see Eq.~(\ref{w17i})), $w_{25-28}$ (see
Eq.~(\ref{w2528})), and $w_{33,34}$ (see Eq.~(\ref{w3334a})) are
simpler. In fact this is the value used in the original BSSN
formulation of GR. We remark that taking $\xi=0$ corresponds
to the case where the momentum constraint is not added to the
evolution equation for $^3\tilde{\Gamma}^{i}$
(cf. Eq.~(\ref{Gamtilde})), and that choice makes the system not even
weakly hyperbolic since the eigenvalues~(\ref{eigenomega}) become
imaginary. This is a feature that was already present in pure
GR~\cite{Alcubierre08a}. Actually, when the scalar field $\phi$ is
absent, the hyperbolicity analysis presented in this Section is
similar (for $\zeta=4$ and $\xi=2$) to the corresponding analysis of
the standard BSSN formulation~\cite{Alcubierre08a}. 
Finally, it is important to mention that in order to avoid any possible 
divergence in the eigenfields or in the source terms (the terms not contributing 
to the principal part of the evolution equations) of both formulations, one needs to consider only STT with 
$f(\phi)>0$ ({\em i.e.} a positive definite NMC function). This in turn precludes the possibility 
of having an infinite or negative effective gravitational constant $G_{\rm eff}$ (cf. Eqs.~(\ref{Geff})).



\section{Discussion}
\label{sec:discussion}

In this paper we have constructed two novel first order strongly
hyperbolic formulations of the STT in the Jordan frame along the lines
of the BM and BSSN approaches. Such constructions show that both
formulations have a well-posed Cauchy problem.  This analysis fills
the gap of a previous study on the Cauchy problem of
STT~\cite{Salgado06} and confirms that the Jordan frame is
mathematically adequate for treating the initial value problem.

One of the most interesting features of the formulations presented
here is that a modified Bona-Mass\'o slicing condition is required for
the two new systems to be strongly hyperbolic while allowing several
slicings ($f_{BM}>0$) which are natural generalizations of the
slicings used in pure GR. In particular $\Theta=1,f_{BM}$ 
in Eq.~(\ref{STTBMlapse}) are two simple choices that lead to 
well behaved eigenfields.

In the absence of a scalar field, the equations of Sec.IIIA,B reduce
(for $\varsigma=-1$, $\zeta=4$ and $\xi=2$) to the known BM and BSSN
formulations of GR (when the NMC function is trivial,
{\em i.e.}, $F(\phi)\equiv 1$, the ``gravitational'' and the scalar-field 
sectors decouple completely up to principal part).

What remains to be investigated is the usefulness and robustness of
these formulations in actual numerical experiments, as well as the
inclusion of a ``live shift''.  Actually, we plan to analyze the
dynamical transition to the phenomenon of spontaneous scalarization in
boson stars arising in STT \cite{Whinnett00} and the subsequent gravitational collapse
to a black hole with gravitational wave emission of scalar type, using
one or both of the hyperbolic formulations presented here. Both
phenomena (spontaneous scalarization and scalar gravitational waves) 
are even present in spherical symmetry due to the NMC,
therefore by assuming such a symmetry one can simplify the equations
without eliminating the interesting physical features.
\bigskip

An important consequence of the analysis presented here is that a
slightly more general STT which includes a function $\omega(\phi)$ in
the kinetic term of the scalar-field sector of the
action~(\ref{jordan}) ({\em i.e.} the kinetic term has the form
$\omega(\phi)(\nabla \phi)^2/2$) posses a well-posed Cauchy problem as
well (except for some choices of $\omega(\phi)$; see
Ref.~\cite{Faraoni07}).  The fact is that all the terms with
$\omega(\phi)$ do not contribute to the principal part of the
equations associated with the {\em metric}\/ sector ({\em i.e.} the
equivalent of Eq.~(\ref{Einst})), while the scalar-field sector (the
equivalent of Eq.~(\ref{KG})) preserves the quasilinear diagonal
hyperbolic form (see Ref.~\cite{Faraoni07} for the detailed
equations). Thus, up to principal part such STT are identical to the
ones analyzed here. The relevance of this generalization is that such
STT can be mapped to the so-called {\em modified theories of
gravity}\/ which are given by a Lagrangian density $f(R)$ ($R$ being
the Ricci scalar)~\cite{Faraoni07}. Some specific choices of $f(R)$
lead to gravity theories which have been recently analyzed in several
contexts. Notably, in the cosmological setting such theories have been
proposed as an alternative to dark energy, since they can produce an
accelerating expansion of the Universe without any exotic form of
matter~\cite{Carroll04,Capozziello03}. However, it must be emphasized
that some of these theories might violate the solar system
tests~\cite{Erickcek06,Chiba07}, and some modifications are required
to circumvent such drawbacks (see Ref.~\cite{Nojiri07} for a review).  The
point we want to underline here is that, from the mathematical point
of view, the viability of such theories relies heavily on the
well-posedness of the Cauchy problem.


\acknowledgments

This work was supported in part by CONACyT grants
SEP-2004-C01-47209-F, and 149945, by DGAPA-UNAM grants
IN112401 and IN119005. DM acknowledges support from CONACyT. 

\newpage

\section{Appendix}
\label{sec:appendix}

\subsection{Full system of equations}

The full system of evolution equations for the generalized Bona-Mass\'o
formulation of STT is the following,
\begin{eqnarray}
& \frac{d}{dt} Q_i\,= &  D_i \left( \alpha \, \Pi\right) \,\,\,,\\
& \frac{d}{dt} \Pi\,= & \alpha\,\Pi\,K + \alpha\,Q^c\,D_c{\rm ln} \left( \alpha\right) + \alpha\,D_c\,Q^c \nonumber \\
& & - \frac{\alpha}{f\left( 1 + \frac{3\,{f^{\prime}}^2}{2f}\right)} \left[\rule{0mm}{0.5cm}   fV^{\prime} - 2f^{\prime}V \right.  \nonumber \\
& & \left. - \frac{1}{2}f^{\prime}\left( 1+3f^{\prime\prime}\right)\left( Q^2 - \Pi^2\right)  
+ \frac{1}{2}f^{\prime} T_{\rm matt}  \rule{0mm}{0.4cm}\right], \,\,\,\nonumber \\
\\
& \frac{d}{dt} \alpha\,= & -\, \alpha^2\, \left( f_{BM}\,K - \Theta \frac{f^\prime}{f} \,\Pi \right) , \\
& \frac{d}{dt} a_i\,= & -\,\partial_i\,\left[ \alpha\, \left( f_{BM}\,K - \Theta \frac{f^\prime}{f} \,\Pi \right)\right] , \\ 
& \frac{d}{dt} {\gamma}_{ij}\,=& -2 \alpha\,K_{ij} \,\,\,, \\
& \partial_0 {d}_{ijk}\,=& -\,\partial_i\left( \alpha\,K_{jk}\right) + \gamma_{l\,(j}\,\partial^2_{k)i}\beta^{l} + d_{ljk}\,\partial_{i}\beta^{l} \nonumber \\
& & + 2 d_{il(j}\,\partial_{k)}\beta^{l}\,\,\,,\\
& \frac{d}{dt} K_{ij}\,= & -D_i D_j\,\alpha + \alpha\,\,^3R_{ij} + \alpha\,K\,K_{ij} - 2\alpha\,K_{il}\,K^{l}_{\,\,j} \nonumber \\
& & + 4\pi G_0\,\alpha \left[\rule{0mm}{0.4cm}\gamma_{ij} \left( S - E\right) - 2\,S_{ij}\right]\,\,\,,\\
&\partial_0 V^{i}\,= & - V^{l}\,\partial_{l}\beta^{i}+
(1+\varsigma) D_{l}\left[\rule{0mm}{0.4cm} \alpha \left( K^{il}-\gamma^{il}K\right) \right] \nonumber \\
&& + \frac{1}{2} \left( \partial^{i}\partial_{m}\beta^{m} -\partial^{m}\partial_{m}\beta^{i}\right) \nonumber \\
& & -\varsigma \alpha \left[\rule{0mm}{0.4cm} a_j \left( K^{ij}-\gamma^{ij}K\right) + 8\pi G_0  J^i\right] \nonumber \\
&& + \alpha\,K^{jk}\,\left(\rule{0mm}{0.4cm}\delta^i_{\,\,j} \,^3\Gamma^{l}_{\,\,kl} -\,^3\Gamma^{i}_{\,\,jk} \right)  \,\,\,.
\end{eqnarray}
where we remind the notations $d/dt:= \partial_t - {\cal L}_{\mbox{\boldmath{$\beta$}}}$ and 
$\partial_0:= \partial_t - \beta^l\partial_l$.

In fact this last equation can also be written as
\begin{eqnarray}
\label{evolV_i}
& \partial_0 V_{i}\,= &  V_{l}\,\partial_{i}\beta^{l}+
\alpha (1+\varsigma) \partial_{l}\left(\rule{0mm}{0.4cm} K^{l}_{\,\,i}-\delta^{l}_{\,\,i}K\right) 
\nonumber \\
&& \!\!\!\!\!\!\!\!
+\alpha a_j \left( K^{j}_{\,\,i}-\delta^{j}_{\,\,i}K\right) 
+ \frac{1}{2} \left( \partial^2_{mi}\beta^{m} -\gamma_{ij}\gamma^{ml}\partial^2_{ml}\beta^{j}\right)
\nonumber \\
& & \!\!\!\!\!\!\!\!
- \alpha\,K^{jk}\,\left\{\rule{0mm}{0.5cm} \gamma_{ij} \left[\rule{0mm}{0.4cm} 2V_k-(2+\varsigma)\,^3\Gamma^{l}_{\,\,kl}\right]
\right. \nonumber \\
&& \!\!\!\!\!\!\!\!
\left.
 +\,^3\Gamma_{ijk} + (1+\varsigma)\,^3\Gamma_{kij} \rule{0mm}{0.5cm}\right\} -\varsigma \alpha 8\pi G_0 J_i
\,\,\,,
\end{eqnarray}
where we note that $\,^3\Gamma^{i}_{\,\,jk} =
\gamma^{il}\left(2d_{(jk)l}-d_{ljk}\right)$ and
$^3\Gamma^{l}_{\,\,kl}= \gamma^{jl}d_{kjl}=\partial_k({\rm
ln}\sqrt{\gamma})$.  So for the optimal choice $\varsigma=-1$,
Eq.~(\ref{evolV_i}) reduces to the usual equation of the BMSS
formulation (cf. Eq.~(8) of Ref.~\cite{Arbona99}).
\bigskip

We also write here the following terms which contain the contributions
of the scalar and matter fields \cite{Salgado06}:
\begin{eqnarray}
\label{Sij}
&& 4\pi G_0 \alpha \left[\rule{0mm}{0.4cm} (S-E)\gamma_{ij} - 2\,S_{ij}\right] =    \nonumber\\
&& -\frac{\alpha}{f}\left[\rule{0mm}{0.4cm}  Q_i\,Q_j (1+f^{\prime\,\prime})+
f^{\prime}(D_iQ_j+\Pi\,K_{ij}) + S^{\rm matt}_{ij} \right]  \nonumber \\
&& + \frac{\gamma_{ij}\alpha}{2f\left( 1+\frac{3{f^{\prime}}^2}{2f}\right)} \left[  \left( Q^2 - \Pi^2 \right) \left( \frac{{f^{\prime}}^2}{2f} - f^{\prime\,\prime}\right) - f^{\prime}V^{\prime} \right. \nonumber\\
&& \left.  - 2V \left( 1+\frac{{f^{\prime}}^2}{2f} \right) + (S_{\rm matt} - E_{\rm matt}) \left( 1+\frac{{f^{\prime}}^2}{f}\right) \right]  \,\,\,,\nonumber \\
\end{eqnarray}
\begin{eqnarray}
&& 4\pi G_0 \alpha \left( S +E\right) = \alpha\,\frac{f^{\prime}}{f}\left( D_m\,Q^m + \Pi\,K \right)    \nonumber\\
&& + \frac{\alpha}{f\left( 1+\frac{3{f^{\prime}}^2}{2f}\right)} \left\lbrace  \Pi^2 \left( 1 + \frac{3{f^{\prime}}^{2}}{4f} + \frac{3\,f^{\prime\prime}}{2} \right) \right. \nonumber \\
&& \left. + Q^2 \left[  \frac{3{f^{\prime}}^{2}}{4f}\left( 1+2f^{\prime\prime}\right) - \frac{f^{\prime \prime}}{2}\right]  \right\rbrace   \nonumber\\
&& + \frac{\alpha}{2f\left( 1+\frac{3{f^{\prime}}^2}{2f}\right)} \left\lbrace S_{\rm matt} + E_{\rm matt} \left( 1 + \frac{3{f^{\prime}}^{2}}{f} \right) \right. \nonumber \\
&& \left.
-2V\left( 1 - \frac{3{f^{\prime}}^{2}}{2f} \right)-3f^{\prime}V^{\prime} \right\rbrace 
\,\,\,,
\end{eqnarray}
\begin{eqnarray}
\label{J_i}
8\pi G_0 J_i =& \frac{1}{f} \left[\rule{0mm}{0.5cm} -f^{\prime}\left(\rule{0mm}{0.4cm} K_{mi}\,Q^m + D_i\Pi\right) \nonumber \right. \\
 & \left. - \Pi\,Q_i \left(\rule{0mm}{0.4cm} 1 + f^{\prime\,\prime} \right)+ J^{\rm matt}_i \rule{0mm}{0.4cm} \right] \,\,\,. 
\end{eqnarray}
The 3-Ricci tensor has the explicit form:
\begin{eqnarray}
\label{3Ricci}
& \!\!\!\!
^3R_{ij} & =\partial_l\, ^3\,\!{\Gamma^l}_{ij}-\partial_i\,^3\,\!{\Gamma^l}_{jl}+
\,^3\,\!{\Gamma^m}_{ij}\, ^3\,\!{\Gamma^l}_{ml} - \,^3\,\!{\Gamma^l}_{im}\,^3\,\!{\Gamma^m}_{lj}
\nonumber \\
& &= -\,\partial_m\, {d^{m}}_{ij} - \partial_{(i}\,\left( 2V_{j)} - {d_{j)m}}^{m}\right) \nonumber \\
& &\,\,\,-\, 4\,{d^{m}}_{mk}\, {d_{(ij)}}^{k}\, + 4 {d^l}_{m\,(i}\,{d_{j)l}}^{m}
-{d_{il}}^{m} {d_{jm}}^{l} \nonumber\\
& & \,\,\, + \,2{d^l}_{mi}\left({d_{lj}}^{m} - {d^{m}}_{lj}\right) 
+ {d_{lm}}^{m}\left(2{d_{(ij)}}^{l}-{d^{l}}_{(ij)}\right) \,\,\,.\nonumber \\
\end{eqnarray}
This expression can be easily rewritten in the form used by the BMSS
formulation (cf. Eq.~(10) of Ref. \cite{Arbona99}).
\bigskip

On the other hand, the full system of evolution equations for the generalized
BSSN formulation of STT is the following:
\begin{eqnarray}
& \frac{d}{dt} Q_i\,= & D_i \left( \alpha \, \Pi\right) \,\,\,,\\
&\frac{d}{dt} \Pi\,= & \alpha\,\Pi\,K + \alpha\,Q^c\,D_c{\rm ln} \left( \alpha\right) + \alpha\,D_c\,Q^c \, \nonumber \\
&  & - \frac{\alpha}{f\left( 1 + \frac{3\,{f^{\prime}}^2}{2f}\right)} \left[\rule{0mm}{0.5cm}  fV^{\prime} - 2f^{\prime}V \right.  \nonumber \\
& & \left. - \frac{1}{2}f^{\prime}\left( 1+3f^{\prime\prime}\right)\left( Q^2 - \Pi^2\right)  + \frac{1}{2}f^{\prime} T_{\rm matt}  \right] \,\,\,,\nonumber \\
 \\
&\frac{d}{dt} \alpha\,= & -\, \alpha^2\, \left( f_{BM}\,K - \Theta \frac{f^\prime}{f} \,\Pi \right) \,\,\,, \\
&\partial_0 \psi = & -\frac{1}{6}\left(\alpha K-\partial_l \beta^l\right) \,\,\,, \\
&\frac{d}{dt} a_i\,= & - \partial_i \left[ \alpha\,\left(f_{BM}\,K-\Theta\frac{f^\prime}{f}\,\Pi\right)\right] \,\,\,, \\ 
&\frac{d}{dt} \tilde{\gamma}_{ij}\,=&-2 \alpha\,e^{\left( \zeta -4 \right) \psi}\,\tilde{A}_{ij} \,\,\,, \\
&\partial_0 \tilde{d}_{ijk}\,= &- \partial_i\,\left( \alpha\, e^{\left( \zeta -4 \right) \psi}\,\tilde{A}_{jk}\right) + \tilde{d}_{ljk}\,\partial_{i}\beta^{l}  \nonumber \\
& & 
+ 2 \tilde{d}_{il(j}\,\partial_{k)}\beta^{l}
-\frac{2}{3}\tilde{d}_{ijk}\partial_l\beta^l - \frac{1}{3}\tilde{\gamma}_{jk}\partial^2_{il}\beta^l \nonumber \\
& & + \tilde{\gamma}_{l\,(j}\,\partial^2_{k)i}\beta^{l}  \,\,\,, \\
\label{Psii}
&\partial_0 \Psi_i\,= & -\frac16\,\partial_i\left( \alpha\,K -\partial_l \beta^l\right) + \Psi_l\partial_i \beta^l\,\,\,, \\
\label{Kbssn}
&\frac{d}{dt} K\, = & -D_iD^i\alpha + \alpha \left(e^{2\left( \zeta -4 \right) \psi} \tilde{A}_{ij}\,\tilde{A}^{ij} + \frac{1}{3}\,K^2\right) \nonumber \\
& & + 4 \pi G_0\,\alpha\left( E + S \right) \,\,\,,  \\
\label{evtildeAij}
&\!\!\!\! \frac{d}{dt} \tilde{A}_{ij}\,\!= &\!\! e^{-\zeta\,\psi}\,\left[\rule{0mm}{0.5cm}\! -D_i\,D_j\,\alpha 
+ \alpha\left(\rule{0mm}{0.4cm}\,^3 R_{ij} - 8\pi G_0\,S_{ij}\right)\right]^{\rm TF} \nonumber \\
& & 
+ \alpha\,\left[\left( \frac{\zeta + 2}{6}\right) \,K\,\tilde{A}_{ij}  - 2\,e^{\left( \zeta -4\right)\psi } \tilde{A}_{ik}\tilde{A}^{k}_{\;j}\right] \,\,\,,\nonumber \\
\end{eqnarray}
where $S_{ij}^{\rm TF}$ can be extracted from Eq.~(\ref{Sij}). In
order to have a full first order formulation it is understood that the
second order spatial derivatives of the lapse and the physical
3-metric appearing in the above equations have to be written
respectively in terms of first derivatives of $a_i$, $\tilde{d}_{ijk}$
and $\Psi_i$.

The evolution equation for the $^3\tilde{\Gamma}^{i}$ is:
\begin{eqnarray}
\label{Gamma^i}
& \!\!\!\! \partial_0\!\,^3\tilde{\Gamma}^{i} = & \!\! \alpha (\xi-2)e^{\left( \zeta - 4\right)\psi}\partial_l \tilde{A}^{il} 
\!-\!\alpha\xi \!\left( \frac{2}{3}\tilde{\gamma}^{il}\partial_l K \!+ 8\pi G_0 \tilde J^i\right) \nonumber \\
&& 
+ 2\alpha e^{\left( \zeta - 4\right)\psi} \,\tilde{A}^{lm}\,\left\{\left[\frac{\xi}{2}\left(\zeta+2\right)
+ 4-\zeta\right]\Psi_l \delta^i_{\,\,m} \right. \nonumber \\
&&
\left. - \delta^i_{\,\,m} a_l + \frac{\xi}{2} {^3\tilde\Gamma^i}_{\,\,lm}\right\} \nonumber \\
&& + \tilde{\gamma}^{jk}\,\partial_{jk}^2\,\beta^i + \frac{1}{3} \tilde{\gamma}^{ij}\,\partial^2_{kj}\,\beta^k - \,^3\tilde{\Gamma}^j\,\partial_j \beta^i \nonumber \\
&& + \frac{2}{3}\,^3\tilde{\Gamma}^i\,\partial_j \beta^j \,\,\,,
\end{eqnarray}
where $\tilde J^i:= \tilde{\gamma}^{il}J_l$ whose explicit expression can be obtained from Eq.~(\ref{J_i}).

We stress that the above equations reduce to the corresponding
equations of the original BSSN formulation for $\zeta=4$ and
$\xi=2$~\footnote{For comparison purposes with the original BSSN
equations, a typo in sign has to be taken into account in the second
line of Eq.~(24) of \cite{Baumgarte:1998te}.}.

We must note that for tensor densities ${\cal D}$ of weight $w$ (here
we omit the rank) the Lie derivative has an extra weight term:
\begin{equation}
\label{Liedens}
{\cal L}_{\mbox{\boldmath{$\beta$}}}\,{\cal D}= \left[{\cal L}_{\mbox{\boldmath{$\beta$}}}\,{\cal D}\right]_{w=0}
+ w {\cal D} \partial_m\beta^{m}\,\,\,,
\end{equation}
where the weight $w$ is related to the power of the determinant of the
physical metric as $\gamma^{w/2}$, and the first term on the right-hand side of
Eq.~(\ref{Liedens}) is to be understood as the ordinary Lie derivative
(as though ${\cal D}$ were a true tensor). For instance, in the case
of $\tilde \gamma_{ij}$ and $\tilde A_{ij}$, we see from
Eqs.~(\ref{tildegamma}) and (\ref{tildeAij}) that these are (0,2)
tensor densities of weight $-2/3$ and $-\zeta/6$, respectively so
\begin{equation}
{\cal L}_{\mbox{\boldmath{$\beta$}}}\,{\cal D}_{ij} = \beta^{m}\partial_m {\cal D}_{ij} + {\cal D}_{im} \partial_j\beta^{m} + {\cal D}_{mj} \partial_i\beta^{m} + w{\cal D}_{ij} \partial_m\beta^{m}\,,
\end{equation}
where ${\cal D}_{ij}$ stands for $\tilde \gamma_{ij}$ and $\tilde A_{ij}$ with 
$w=-2/3$ and $w=-\zeta/6$, respectively.
\bigskip

For completeness, we provide the expression of the 3-Ricci tensor
$^3R_{ij}$ in terms of the conformal metric (\ref{tildegamma})
\begin{equation}
^3 R_{ij} = \,^3\tilde{R}_{ij} + \,^3R^{\psi}_{ij} \,\,\,,
\end{equation}
 where
\begin{eqnarray}
\label{3Ricciconf}
& ^3\tilde{R}_{ij} =& -\frac{1}{2}\, \tilde{\gamma}^{lm}\, \partial^2_{lm}{\tilde{\gamma}}_{ij} + {\tilde{\gamma}}_{k\,(i}\,\partial_{j)}\,\!^3\tilde{\Gamma}^{k} + \,^3\tilde{\Gamma}^k\,^3\tilde{\Gamma}_{(ij)\,k} \nonumber \\
& & + \tilde{\gamma}^{lm}\,\left( 2\,^3\tilde{\Gamma}^{k}_{\,\,\,l\,(i}\,^3\tilde{\Gamma}_{j)\,km} + \,^3\tilde{\Gamma}^{k}_{\,\,\,im}\,^3\tilde{\Gamma}_{klj}\right) , \,\,\, \\
& ^3R^{\psi}_{ij} := & -2\,\tilde{D}_i \tilde{D}_j\,\psi\, -2 \tilde{\gamma}_{ij}\,\tilde{D}^k \tilde{D}_k\,\psi\, + 4(\tilde{D}_i\psi)\,(\tilde{D}_j\psi) \nonumber \\
& & -4\,\tilde{\gamma}_{ij}(\tilde{D}^k\,\psi)\,(\tilde{D}_k\,\psi) \,\,\,,
\end{eqnarray}
where $\tilde D$ stands for the covariant derivative associated with
the conformal metric $\tilde\gamma^{ij}$, and $^3\tilde{\Gamma}_{kij}:=
\tilde{\gamma}_{kl} {^3\tilde{\Gamma}^{l}}_{ij}$. In fact by putting
tildes in Eq.~(\ref{3Ricci}), and noticing that $ {{\tilde
d}_{km}}^{\;\;\;\;m}=\, ^3\tilde\Gamma^{l}_{\,\,kl}= \partial_k({\rm
ln}\sqrt{\tilde\gamma})\equiv 0$ (since $\tilde\gamma:=1$), we see that
Eq.~(\ref{3Ricciconf}) can be written as Eq.~(\ref{3Ricci}) with the
terms involving ${{\tilde d}_{lm}}^{\;\;\;\;m}$ dropped. Moreover, 
the constraint equations (\ref{CEHfSST}) and (\ref{CEMfSST}) can be easily rewritten in terms of 
the quantities defined in Eqs.~(\ref{tildegamma})$-$(\ref{Aij}).

\subsection{Simple example of an hyperbolic system}
Let us consider the system
\begin{eqnarray}
\label{toy1}
 \partial_t u_1 + a\,\partial_x u_1 + b\,\partial_x u_2 = S_{u_1}(u_1,u_2) \,\,\,,\\
\label{toy2}
 \partial_t u_2 + c\,\partial_x u_1 + d\,\partial_x u_2 = S_{u_2} (u_1,u_2) \,\,\,\,,
\end{eqnarray}
where the coefficients $a-d$ are in general functions of $(t,x)$, and
the $S_{u_1,u_2}(u_1,u_2)$ are {\it source} functions.  Using the {\em
judicious guessing}\/ approach we shall construct a new system of the
form~(\ref{PDE3}) which is manifestly hyperbolic, and then we shall
confront this method with the more systematic method which uses linear
algebra.

First we ask for a linear combination $w:=u_1 + \sigma u_2$ to be the
eigenfunction. This means that we write an evolution equation for $w$,
where $\sigma(a,b,c,d)$ is a function of the coefficients $a-d$, such
that the r.h.s takes the form $-\lambda_{\sigma} \, \partial_x w $,
where $\lambda_{\sigma}$ will be the eigenvalue. In this way we have
\begin{eqnarray}
\label{jg1}
&\partial_t \left(u_1 + \sigma u_2\right) &\simeq - \partial_x \left[\rule{0mm}{0.4cm} \left( a+\sigma c\right) u_1 + \left( b + \sigma d\right) u_2 \right] \nonumber \\
& &\simeq -\left( a+\sigma c\right) \,\partial_x\left(  u_1 + \frac{b+\sigma d}{a+\sigma c} u_2\right) \,\,\,\,.\nonumber\\
\end{eqnarray}
Note that with the symbol $\simeq$ we have discarded from the analysis
all the terms which do not contain derivatives of $u_1$ and $u_2$
(including the sources). In this way we can assume as that the
coefficients $a-d$ are ``constants''. By comparing the coefficients of
$u_2$ on both sides of Eq.~(\ref{jg1}), we see that we need to take
$\sigma= (b+\sigma d)/(a+\sigma c)$, which implies $\sigma_{\pm} = \frac{d
- a \pm \sqrt{\left( a-d\right) ^2 + 4bc}}{2c}$. Moreover, the
eigenfunctions will be smooth provided $c(t,x)$ does not vanish in
some spacetime point.

Therefore we conclude that the two eigenfunctions are $w_\pm= u_1 +
\sigma_{\pm} u_2$ with the corresponding eigenvalues
\mbox{$\lambda_{\pm} = a+\sigma_\pm c=\frac{a + d \pm \sqrt{\left(
a-d\right) ^2 + 4bc}}{2}$}.  The eigenvalues are real and the
eigenfunctions are nondegenerate if $\left( a-d\right) ^2 + 4bc >0$
and $c\neq 0$. If this condition holds and the eigenfields are smooth
then the system is strongly hyperbolic
\footnote{If the function $c(t,x)$ is globally null, the equation for
$u_2$ decouples and $u_2$ is itself an eigenfunction propagating with
speed $d$; the other eigenvalue and eigenfunction can be obtained by
proceeding in the same way, taking $c=0$ in Eq.~(\ref{jg1}).  This
results in $w_2= u_1 + bu_2/(a-d)$ (with $a\neq d$) and $\lambda_2=
a$. However, if $c=0$, $a=d$ and $b\neq 0$ then the system degenerates
and becomes weakly hyperbolic. In such a case the characteristic
matrix is a {\em Jordan block} (cf. Eq.~(\ref{toysys})), which implies
that it cannot be diagonalized. A similar situation happens if $b=0$,
$a=d$ and $c\neq 0$. On the other hand, if $c=0=b$ then both
equations~(\ref{toy1}) and~(\ref{toy2}) decouple and therefore $u_1$
and $u_2$ are themselves eigenfunctions (or any linear combination of
them is an eigenfunction as well if in addition $a=d$).}.

Now, let us find the eigenfunctions of the same system using the
standard method of linear algebra. We rewrite the system in a more
convenient matrix form as:
\begin{equation}
\label{toysys}
\partial_t \left( \begin{array}{cc} u_1 \\ u_2 \end{array}\right) + \left( \begin{array}{cc} a &b \\ c &d \end{array}\right) \partial_x \left( \begin{array}{c} u_1 \\ u_2 \end{array}\right) \simeq 0 \,\,\,.
\end{equation}
The eigenvalues $\lambda$ of the characteristic matrix turn out to be
$\lambda_{\pm} = \frac{a + d \pm \sqrt{\left( a-d\right) ^2 +
4bc}}{2}$, which are identical to the ones found before. On the other
hand, the eigenvectors are:
\begin{equation}
v_{+}= \left( \begin{array}{c} -\sigma_- \\ 1 \end{array}\right) 
\,\,\,,\,\,\,v_{-}= \left( \begin{array}{c} \sigma_+  \\ -1 \end{array}\right) \,\,\,.
\end{equation}
Thus, the matrix of eigenvectors reads 
\begin{equation}
 \mathbb{R} = \left( \begin{array}{cc} -\sigma_{-} &\sigma_{+}\\ 1 &-1 \end{array}\right)\,\,\,.
\end{equation}
Then we can write the eigenfunctions as:
\begin{equation}
\left(\begin{array}{c} f_1\\ f_2 \end{array}\right) \!= \mathbb{R}^{-1}\!\left( \begin{array}{c} u_1\\ u_2 \end{array}\right) \! =\! \frac{1}{\sigma_{+} - \sigma_{-}}\! \left( \begin{array}{c} u_1 + \sigma_{+} u_2 \\ u_1 + \sigma_{-} u_2 \end{array}\right) ,
\end{equation}
which are the same found before modulo the factor
\mbox{$\frac{1}{\sigma_{+} - \sigma_{-}}= c/\sqrt{\left( a-d\right) ^2
+ 4bc}$}. The fact that we do not obtain exactly the same
eigenfunctions is a consequence of the fact that the eigenvectors are
unique only up to a rescaling. However, both pairs of eigenfunctions
$w_\pm$ and $f_{1,2}$ are equivalent.

Finally, notice that if $a=d$, then $\lambda_{\pm} = a \pm \sqrt{bc}$,
and $w_\pm= u_1 \pm \sqrt{b/c}\,u_2$. Thus $a$ plays the same roll as
$-\beta^i$ in the evolution equations of Sec.II. For instance, if the
coefficients $b,c$ are null, the equations (\ref{toy1}) and
(\ref{toy2}) decouple and then $u_1$ and $u_2$ are themselves
eigenfunctions, indicating that they propagate along the ``normal 
lines'' (with speed $a$). This is precisely what happens with
eigenfunctions $w_{1-18}$ of Sec.IIIA.  On the other hand, when $b$
and $c$ are non-null, these coefficients are related to the
propagation of the eigenfields some of which propagate along the
``light cones''. For instance, if $b$ is identified with $\alpha
\gamma^{xx}$ (or $-\varsigma \alpha \gamma^{xx}$, or even $\alpha
f_{BM} \gamma^{xx}$) and $c$ with $\alpha$, in many of the evolution
equations of Sec.IIIA (assuming propagation only in the `$x$'
direction), then the $\lambda_{\pm}$ are in turn to be identified with
the speeds of propagation $-\beta^x \pm \alpha \sqrt{\gamma^{xx}}$ or
$-\beta^x \pm \alpha \sqrt{-\varsigma\gamma^{xx}}$ or even $-\beta^x
\pm \alpha \sqrt{f_{BM}\gamma^{xx}}$ (cf. Eqs.~(\ref{l23-28}),
(\ref{l29-32}) and (\ref{l3334}), respectively) of several
eigenfields ({\em e.g.} $w_\pm= u_1 \pm \sqrt{\gamma^{xx}} u_2$, or $w_\pm=
u_1 \pm \sqrt{-\varsigma\gamma^{xx}} u_2$, or even $w_\pm= u_1 \pm
\sqrt{f_{BM}\gamma^{xx}} u_2$, where $u_1$ and $u_2$ represent field
variables like those entering in the expressions for $w_{23-28}$,
$w_{29-32}$ and $w_{33,34}$ respectively).


\bibliographystyle{bibtex/apsrev}
\bibliography{bibtex/referencias}


\end{document}